\documentclass{article}
\usepackage{geometry}[margins = 1.5 cm]
\usepackage{graphicx}
\usepackage{newtxtext}
\usepackage{newtxmath}
\usepackage{natbib}
\usepackage{hyperref}
\usepackage{gensymb}
\hypersetup{
    colorlinks = true,
    urlcolor   = blue,
    citecolor  = black,
}
\usepackage{lscape}
\usepackage{authblk}
\usepackage{cleveref} 
\usepackage{amsmath}

\newcommand{\RomanNumeralCaps}[1]
\linenumbers

\DeclareMathOperator*{\argmin}{arg\,min}
\newcommand{\ON}{\text{ON}}
\newcommand{\OFF}{\text{OFF}}
\newcommand{\derp}[2]{\frac{\partial #1}{\partial #2}}
\newcommand{\der}[2]{\frac{d #1}{d #2}}
\newcommand{\E}{\mathbb{E}}
\newcommand{\R}{\mathbb{R}}

\newcommand{\Pran}{\textit{Pr}}
\newcommand{\Ray}{\textit{Ra}}
\newcommand{\Lew}{\textit{Le}}
\newcommand{\Asp}{\textit{A}}

\newcommand{\upi}{\pi}
\newcommand{\thalf}{\frac{1}{2}}

\newcommand{\bph}{\boldsymbol{\phi}}
\newcommand{\bt}{\boldsymbol{\theta}}

\newcommand{\rom}[1]{\uppercase\expandafter{\romannumeral #1\relax}}

\title{Most Likely Noise-Induced Overturning Circulation Collapse in a 2D Boussinesq Fluid Model}

\author[1]{Jelle Soons \footnote{Correspondence: j.soons@uu.nl}}
\author[2]{Tobias Grafke}
\author[1]{Henk A. Dijkstra}

\affil[1]{Institute for Marine and Atmospheric research Utrecht, Utrecht University, Princetonplein 5, 3584 CC Utrecht, The Netherlands}
\affil[2]{Mathematics Institute, University of Warwick, Coventry CV4 7AL, United Kingdom}

\begin{document}
\maketitle

\begin{abstract}
There is a reasonable possibility  that the present-day Atlantic Meridional Overturning Circulation  is in a bi-stable regime 
and hence it is relevant to compute probabilities and pathways of noise-induced transitions between the stable equilibrium states. 
Here, the most probable transition pathway of a noise-induced collapse of the northern overturning circulation  in a  spatially-continuous 
two-dimensional  model with surface temperature and stochastic salinity forcings  is directly computed using Large Deviation Theory (LDT). 
This pathway reveals the fluid dynamical  mechanisms of such a  collapse. Paradoxically it starts off with a strengthening of the  
northern overturning circulation before a short but strong salinity pulse induces a second overturning cell. The increased atmospheric 
energy input  of this two-cell configuration  cannot be mixed  away quickly enough,  leading to the collapse of the northern  
overturning cell and finally  resulting in a southern overturning circulation. Additionally, the approach allows us to compare the probability of this collapse under different parameters in the deterministic part of the salinity surface forcing, which quantifies the increase in collapse 
probability as the bifurcation point of the system is approached. 
\end{abstract}


\section{Introduction}
\label{sec:introduction}
The Atlantic Meridional Overturning Circulation (AMOC) plays a vital role in regulating Earth's climate. It transports warm upper ocean water northward which causes the rather mild climate in Western Europe. When reaching the subpolar North Atlantic Ocean the relatively warm and salty water is cooled by the atmosphere and becomes the denser cold and salty North Atlantic Deep Water (NADW), which sinks and returns as a deep southward flow \citep{frajka2019atlantic}. The AMOC is identified as a tipping element in the present-day climate system \citep{armstrong2022exceeding}, and its collapse would have severe consequences on the global climate \citep{van2024physics}.

\citet{stommel1961thermohaline} already realized the AMOC's potential for tipping, as the AMOC is affected by two competing feedbacks due to the opposing effects of temperature and salinity on the density of seawater. A strengthening of the AMOC results in increased northward heat transport, causing a decrease in density of the upper subpolar North Atlantic water which inhibits sinking. Consequently, the thermal circulation represents a negative feedback. On the other hand, AMOC strengthening also increases northward salt transport which aides the northern deep water formation. Hence the haline circulation provides a positive feedback mechanism: the salt-advection feedback \citep{marotzke2000abrupt}. These feedbacks, together with the fact that ocean temperature anomalies are more strongly damped by the atmosphere than ocean salinity anomalies, result in a multiple equilibria regime for the AMOC. In the box model used by \citet{stommel1961thermohaline} two stable equilibria exist in a range of the surface buoyancy forcing, and hence transitions between these states can occur. 

Since then multiple equilibria regimes have been found in a hierarchy of AMOC models, from conceptual box models \citep{rooth1982hydrology, rahmstorf1996freshwater, lucarini2005thermohaline, cimatoribus2014meridional}, two-dimensional (2D) models \citep{quon1992multiple, thual1992catastrophe},  three-dimensional (3D) ocean-only models \citep{dijkstra2007characterization}, various Earth system Models of Intermediate Complexity (EMICs) \citep{rahmstorf2005thermohaline}, Global Climate Models (GCMs) \citep{hawkins2011bistability}, to  modern Earth System Models \citep{van2023asymmetry}.  Based on observations of available stability indicators, such as the AMOC induced freshwater convergence, \citep{dijkstra2007characterization}, the present-day AMOC would be in such regime \citep{weijer2019stability}, although this claim is contested \citep{jackson2018hysteresis}. Accordingly there has been a growing interest in establishing the distance of the present-day AMOC to these bifurcation points, and whether it will cross a critical threshold in a bifurcation-induced tipping event. Based on early warning signals determined from reconstructed historical data the AMOC is thought to be heading towards this critical bifurcation point \citep{caesar2018observed, boers2021observation}, and estimates have been made for when this point will be reached \citep{ditlevsen2023warning}. However, still many uncertainties, in particular in the 
sea surface temperature data to reconstruct the historical AMOC,  remain \citep{ben2024uncertainties}.

In addition to a bifurcation-induced tipping the AMOC can also undergo a noise-induced tipping, where the transition occurs due to variations in small-scale forcing processes represented as `noise' \citep{dijkstra2024role}. This is far more dangerous as there are no reliable early-warning signals for a noise-induced transition as opposed to a bifurcation-induced one \citep{ditlevsen2010tipping}. Moreover, a bifurcation-induced tipping can only occur close to the bifurcation thresholds, whereas one induced by noise can occur as long as the system is in a multiple-equilibria regime. Several AMOC models have been developed where the ocean's surface is forced by stochastic salinity noise in order to study the statistics of these noise-induced transitions \citep{cessi1994simple, timmermann2000noise}. A major hurdle in this analysis is that noise-induced AMOC transitions are expected to be quite rare. There is no observational evidence for such a transition over the historical period, and computed probabilities for a transition in box models 
can be as low as $10^{-8}$ under realistic noise levels \citep{castellana2019transition, van2024role}. At these low probabilities standard Monte-Carlo techniques fail to obtain a transition path within reasonable computing time. However, we would still like to obtain these rare transition paths as they can be qualitatively different from the more commonly studied bifurcation-induced transitions \citep{soons2023optimal} and they may provide new early-warning signals for a noise-induced transition \citep{giorgini2020precursors}. 

Here, we use a technique from the Freidlin-Wentzell Theory of Large Deviations \citep{freidlin1998random} to directly compute the most likely transition path in the low-noise limit of a noise-induced overturning circulation collapse in a 2D Boussinesq fluid model with stochastic surface salinity forcing. 
This is achieved by minimizing the Freidlin-Wentzell action yielding this most likely transition path and its associated noise forcing. This minimization is equivalent to maximizing the probability of the applied stochastic forcing under the constraints that it causes a transition. This technique transforms this rare transition sampling problem into a deterministic minimization problem. The corresponding path, also called the instanton, has been determined  in other applications to study noise-induced transitions \citep{grafke2015instanton, grafke2019numerical, woillez2020instantons, schorlepp2022spontaneous}, and we have previously computed the instantons for an AMOC collapse in a stochastic version of the \citet{wood2019observable} box model \citep{soons2023optimal}. This work is a natural extension upon that.

In the following, we introduce the 2D Boussinesq model of thermohaline flow in section \ref{sec:model}. Then, in section \ref{sec:method} the 
methodology is shortly described, including  a short introduction of the Freidlin-Wentzell Theory of Large Deviations, which is subsequently applied 
to the model. The fluid dynamical mechanisms and energetics of the resulting most-likely overturning circulation collapse are 
analyzed  in section \ref{sec:collapse}. Next, the probability ratios of collapse in the low-noise limit are presented  for 
various surface forcings in section \ref{sec:ratios}. Section \ref{sec:sum} finally contains a summary  and  discussion of 
the results. 

\section{The 2D Boussinesq Model}
\label{sec:model}
We use a Boussinesq model of the zonally averaged thermohaline circulation in the double-hemispheric Atlantic basin with heat and 
salt  forcing at the surface, where the latter has a stochastic component. It is considered for a rectangular basin of length $L$ and depth $H$. The 
diffusivities of heat $\kappa_T$, salt $\kappa_S$, and momentum $\nu$ are assumed constant and must be interpreted as eddy 
diffusivities. A linear equation of state holds with thermal and haline compressibility coefficients $\alpha_T$ and $\alpha_S$, respectively. 
The model used is a slight modification of the ones studied in \citet{quon1992multiple}, \citet{thual1992catastrophe} and 
\citet{dijkstra1997symmetry}.

The governing equations are non-dimensionalised with scales $H$, $H^2/\kappa_T$, $\kappa_T/H$, $\Delta T$ and $\Delta S/\lambda$ for length, time, velocity, temperature, and salinity, respectively. Here, $\Delta T$ and $\Delta S$ are characteristic meridional temperature and salinity differences, and $\lambda$ is the buoyancy ratio $\left(\lambda = (\alpha_S\Delta S)/(\alpha_T\Delta T)\right)$. The streamfunction $\psi$ and vorticity $\omega$
are introduced with the  horizontal and vertical velocity $(u,w)$ expressed by $u = \partial_z\psi$, $w = -\partial_x\psi$ and $\omega = \partial_xw-\partial_zu$. Hence the system is described by streamfunction $\psi(x,z,t)$, vorticity $\omega(x,z,t)$, temperature $T(x,z,t)$ and salinity $S(x,z,t)$. The non-dimensional governing equations are 
 \begin{eqnarray}
  \Pran^{-1}\left(\derp{\omega}{t}+\derp{\psi}{z}\derp{\omega}{x}-\derp{\psi}{x}\derp{\omega}{z}\right) & = & \nabla^2\omega + \Ray\left(\derp{T}{x}-\derp{S}{x}\right), \label{eq:momentum}\\
  \omega & = & -\nabla^2\psi, \label{eq:stream}\\
  \derp{T}{t} + \derp{\psi}{z}\derp{T}{x} - \derp{\psi}{x}\derp{T}{z} & = & \nabla^2T + \frac{h(z)}{\tau_T}\left(T_S(x)-T\right), \label{eq:temp}\\
  \derp{S}{t} + \derp{\psi}{z}\derp{S}{x} - \derp{\psi}{x}\derp{S}{z} & = & \Lew^{-1}\nabla^2S + \frac{h(z)}{\tau_S}\left(S_S(x)+\Tilde{S}_S(x,t)\right),\label{eq:salt}
\end{eqnarray}
on the domain $(x,z)\in [0,\Asp]\times[0,1]$ for time $t\geq 0$. The basin's bottom is at $z=0$, its surface at $z = 1$, and the southern and northern end are located at $x = 0$ and $x = \Asp$ respectively. All the boundaries are assumed to be stress free
\begin{align*}
    x = 0,\,\Asp:\quad&\psi = \omega = 0, \\
    z = 0,\,1:\quad&\psi = \omega = 0, 
\end{align*}
and additionally they are isolated and impervious to salt and temperature
\begin{align*}
    x = 0,\,\Asp:\quad&\derp{T}{x} = \derp{S}{x} = 0, \\
    z = 0,\,1:\quad&\derp{T}{z} = \derp{S}{z} = 0.
\end{align*}

Equations \eqref{eq:momentum} and \eqref{eq:stream} model the advection and diffusion of momentum, which is forced by the meridional temperature and salinity gradients. Note that using the streamfunction implies mass conservation. The equations \eqref{eq:temp} and \eqref{eq:salt} represent the advection-diffusion processes for heat and salinity respectively. Temperature is forced by a restoring force, where it is always driven towards a fixed atmospheric temperature $T_S(x)$. Salinity on the other hand is forced by a flux, consisting of a deterministic and temporally constant part $S_S(x)$ and a stochastic part $\Tilde{S}_S(x,t)$. All forcings have a vertical profile prescribed by
\begin{equation*}
    h(z) = \exp\left(\frac{z-1}{\delta_V}\right), 
\end{equation*}
with $\delta_V \ll 1$ a characteristic thickness for the top boundary layer. This way the forcings are essentially only applied near the surface where $h(1) = 1$. Moreover, the temperature and salinity forcings have respective time scales $\tau_T$ and $\tau_S$. 

Lastly, the following non-dimensional parameters are used: the Prandtl number $\Pran$, the Lewis number $\Lew$, the thermal Rayleigh number $\Ray$ and the aspect ratio $\Asp$. These are defined as
\begin{equation*}
    \Pran = \frac{\nu}{\kappa_T},\qquad\Lew = \frac{\kappa_T}{\kappa_S},\qquad\Ray = \frac{g\alpha_T\Delta TH^3}{\nu\kappa_T},\qquad\Asp = \frac{L}{H}, 
\end{equation*}
where $g$ is the standard acceleration of gravity. Throughout this work we take these to be constant: $\Pran = 1$, $\Lew = 1$, $\Ray = 4\cdot10^4$, and $\Asp = 5$ which are based on the values found in \citet{dijkstra1997symmetry}. Moreover, we also fix $\tau_T = 0.1$, $\tau_S = 1$, and $\delta_V = 0.05$.

\subsection{Surface forcing}
The prescribed atmospheric temperature is
\begin{equation*}
    T_S(x) = \frac{1}{2}\left(\cos\left(2\upi \left(\frac{x}{\Asp}-\frac{1}{2}\right)\right)+1\right)
\end{equation*}
which is identical to \citet{dijkstra1997symmetry} and is symmetric around the equator ($x = \thalf$)  with hot tropics and cold poles. Regarding the freshwater forcing an additional constraint is present, as we want the total salinity in the basin to be conserved. Integrating \eqref{eq:salt} over the complete basin yields
\begin{align*}
    \der{}{t}\int_0^1\int_0^\Asp S\,dx\,dz = \frac{\delta}{\tau_S}\left(1-e^{-1/\delta}\right)\left(\int_0^\Asp S_S(x)\,dx + \int_0^\Asp \Tilde{S}_S(x,t)\,dx\right)
\end{align*}
and so total salinity is only conserved if each of the integrated horizontal profiles of the deterministic and stochastic part of the freshwater forcing is zero. For the deterministic part we take
\begin{equation*}
    S_S(x) = 3.5\cos\left(2\upi\left(\frac{x}{\Asp}-\frac{1}{2}\right)\right) - \beta\sin\left(\upi\left(\frac{x}{\Asp}-\frac{1}{2}\right)\right)
\end{equation*}
which is asymmetric for $\beta > 0$ with the northern (sub)polar region being more freshened than its southern counterpart  (and vice versa for 
$\beta < 0$). Note that integrated over the basin's horizontal dimension the freshwater forcing is zero.

For the stochastic component we assume that the noise is  white in time as this simplifies the computations significantly. We choose
\begin{equation*}
    \Tilde{S}_S(x,t) = \sqrt{\frac{\varepsilon}{K}}\sum_{k=1}^K\dot{W}_k^{(1)}(t)\cos\left(\frac{2\upi}{\Asp}kx\right) + \dot{W}_k^{(2)}(t)\sin\left(\frac{2\upi}{\Asp}kx\right)
\end{equation*}
with $2K$ spatial components, variance $\varepsilon>0$ and $2K$ independent Wiener processes $W_k^{(1)}(t)$ and $W_k^{(2)}(t)$ for $k\in\{1,\dots,K\}$. This fulfills the salinity conservation constraint and is white in time as
\begin{align*}
    \E\left(\Tilde{S}_S(x,t)\Tilde{S}_S(x,t')\right) = \varepsilon\delta(t-t') 
\end{align*}
where $\delta(\cdot)$ is the Dirac delta distribution. The number of components $K$ influences the noise behaviour greatly and so a considered choice for its value is required. Of course a larger $K$ will result in a higher resolution in the form of allowing the forcing to vary on smaller spatial scales but it will also require additional CPU-time to compute realizations and the instanton trajectory.  Following \citet{adler2003status, adler2017global, boot2024} the largest variation in mean annual precipitation is around the Inter-Tropical Convergence Zone (ITCZ). This band of high variance stretches for roughly $20\degree$ in latitude. Taking the Atlantic basin from $70\degree$S to $70\degree$N then results in a rough cut-off at $K = 7$, which we fix here in this paper. 

\subsection{Model as SPDE}
The model is rewritten as a standard Stochastic Partial Differential Equation (SPDE) in order to apply the Freidlin-Wentzell Theory to it. We define the function 
\begin{equation*}
    \bph:\,\Omega\times\R_{\geq0}\to\R^3\qquad\mbox{with}\qquad \bph(x,z,t) = \left(\omega(x,z,t),T(x,z,t),S(x,z,t)\right)^T
\end{equation*}
where $\Omega = [0,\Asp]\times[0,1]$ is the basin domain. Note that the streamfunction $\psi(x,z,t)$ is omitted as it has no time evolution equation, and we rewrite it as $\psi[\omega(x,z,t)]$. This is possible since the Poisson equation \eqref{eq:stream} with homogeneous Dirichlet boundaries has a unique well-defined $\psi(x,z,t)$ for every $\omega(x,z,t)$. So we write
\begin{equation*}
    \psi[\omega] = \mathcal{G}(\omega)
\end{equation*}
where $\mathcal{G}$ is the linear operator
\begin{equation*}
    \mathcal{G}(v) = -\int_\Omega G(x,z;\zeta,\xi)v(\zeta,\xi,t)\,d\zeta\,d\xi
\end{equation*}
with $G$ Green's function obeying
\begin{align*}
    \nabla^2_{(x,z)}G(x,z;\zeta,\xi) &= \delta(\zeta-x,\xi-z)\quad\text{ for }(x,z)\in\Omega\\
    G(x,z;\zeta,\xi) &= 0\hspace{2.35cm}\text{ for }(x,z)\in\partial\Omega.
\end{align*}
Now, the deterministic part of the evolution equations is given by 
\begin{equation*}
    f:\,\bph\mapsto\left(f_1[\bph], f_2[\bph], f_3[\bph]\right)^T
\end{equation*}
where
\begin{align*}
    f_1[\bph] &= \partial_x\psi[\phi_1]\partial_z\phi_1-\partial_z\psi[\phi_1]\partial_x\phi_1 + \Pran\nabla^2\phi_1 + \Pran\Ray\left(\partial_x\phi_2-\partial_x\phi_3\right)\\
    f_2[\bph] &= \partial_x\psi[\phi_1]\partial_z\phi_2-\partial_z\psi[\phi_1]\partial_x\phi_2 + \nabla^2\phi_2 + \frac{h(z)}{\tau_T}\left(T_S(x) - \phi_2\right)\\
    f_3[\bph] &= \partial_x\psi[\phi_1]\partial_z\phi_3-\partial_z\psi[\phi_1]\partial_x\phi_3 + \Lew^{-1}\nabla^2\phi_3 + \frac{h(z)}{\tau_S}S_S(x).
\end{align*}

Let $\boldsymbol{\eta}(x,z,t)$ represent the  three-dimensional spatio-temporal white noise with
\begin{equation*}
    \E\left(\eta_i(x,z,,t)\eta_i(x',z',t')\right) = \delta(x-x')\delta(z-z')\delta(t-t')\qquad\text{ for }i\in\{1,2,3\}
\end{equation*}
and note that
\begin{align*}
    &\Big\{\cos\Big(\frac{2\upi}{\Asp}kx\Big)\cos\Big(2\upi lz\Big),\,\cos\Big(\frac{2\upi}{\Asp}kx\Big)\sin\Big(2\pi lz\Big),\,\sin\Big(\frac{2\upi}{\Asp}kx\Big)\cos\Big(2\pi lz\Big),\\
    &\quad\sin\Big(\frac{2\upi}{\Asp}kx\Big)\sin\Big(2\upi lz\Big)\Big\}_{k,l\in\mathbb{N}_0}
\end{align*}
is an orthonormal basis of $L^2\big(\Omega\big)$. We denote the associated basis transform operator with $\mathcal{U}$. As this is a unitary operator we have that its adjoint is equal to its inverse $(\mathcal{U}^* = \mathcal{U}^{-1})$. Furthermore we define projection operator $\mathcal{P}$ which projects a function in $L^2\big(\Omega\big)$ onto the space spanned by this basis under restrictions that $l = 0$ and $1\leq k \leq K$. Then it holds that
\begin{equation*}
    \Tilde{S}_S(x) = \sqrt{\frac{\varepsilon}{K}}\mathcal{U}^*\mathcal{P}\mathcal{U}(\boldsymbol{\eta})(x,z,t).
\end{equation*}

The model can be written then as the SPDE system: 
\begin{equation}
    \partial_t\bph(x,z,t) = f[\bph(x,z,t)] + \sqrt{\varepsilon}\sigma(\boldsymbol{\eta}(x,z,t)),
\end{equation}\label{eq:SPDE}
where the operator $\sigma$ acts on the white-in-space and white-in-time noise
\begin{align*}
    \sigma = \begin{pmatrix}
        0\\
        0\\
        \frac{h(z)}{\tau_S\sqrt{K}}\mathcal{U}^*\mathcal{P}\mathcal{U}
    \end{pmatrix}.
\end{align*}

\subsection{Deterministic Equilibria}
\begin{figure}
    \centering
    \includegraphics[width=\textwidth]{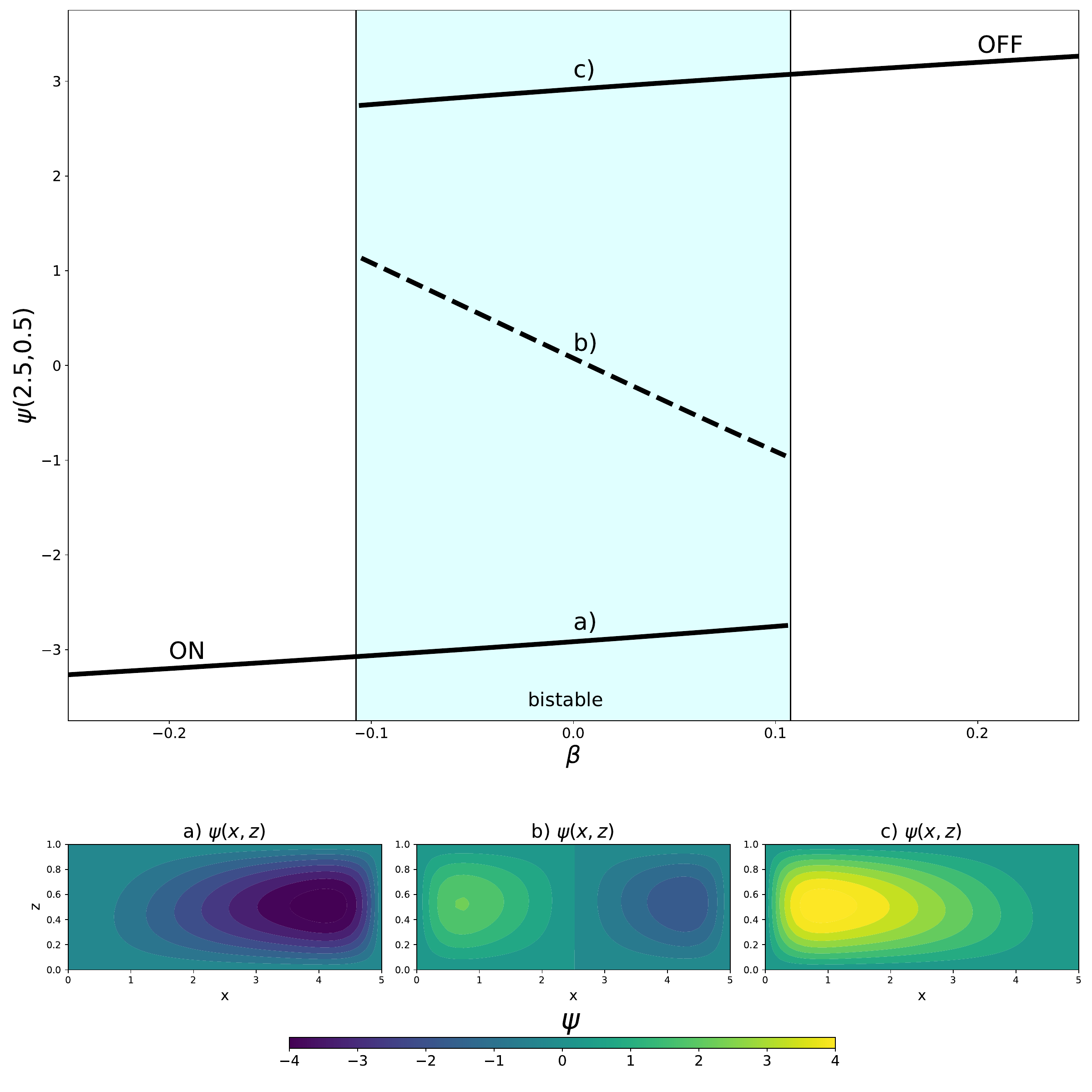}
    \caption{Partial bifurcation diagram of the 2D Boussinesq model (top) with stable ON and OFF branches (solid) and unstable saddle states in the bistable region (dashed), with contour plots of the streamfunction $\psi$ of examples of the stable ON state (a), saddle (b) and stable OFF state (c) which are all indicated on the bifurcation diagram.}
    \label{fig:bifdiagram}
\end{figure}

Under the stated surface forcings, there are three distinct equilibria in the deterministic setting (i.e. $\varepsilon = 0$) that are relevant and they are depicted in the partial bifurcation diagram in figure \ref{fig:bifdiagram}. Here the asymmetry $\beta$ of the deterministic freshwater forcing is the bifurcation parameter. There are two possible stable states which both consist of one asymmetric overturning cell with downwelling near one pole, see figure \ref{fig:bifdiagram}a and c. One state with downwelling near the basin's northern boundary, denoted as ON state (as it mimics an active AMOC state), and one with downwelling in the south, denoted as OFF state (as it mimics a collapsed AMOC state). As parameter $\beta$ increases, the freshwater flux into the northern half increases as well which inhibits downwelling there and weakens the ON state. Eventually, for $\beta > 0.11$ the freshwater flux is large enough to collapse the ON state. Note that the model's dynamics are invariant under the transformations $\beta \to -\beta$ and $x \to \Asp-x$. In other words, the ON state at $\beta$ is exactly the OFF state at $-\beta$ but mirrored at the equator, $x = \Asp/2$. Therefore the OFF state collapses for values $\beta<-0.11$. This creates a bistable regime for $|\beta|<0.11$. Additionally, in this bistable regime a saddle state is present, which is an unstable equilibrium with two symmetric overturning cells in each hemisphere with downwelling near both poles, see figure \ref{fig:bifdiagram}b. We restrict ourselves here to the bistable regime as this contains the relevant attractive structure for the noise-induced collapse. The bifurcation structure outside of this region is in principle much more complicated and a thorough bifurcation analysis and descriptions of the instability mechanisms can be found in \citet{dijkstra1997symmetry}.

The ON and OFF state (also denoted as $\bph_\ON$ and $\bph_\OFF$) are pole-to-pole circulations that each can be qualitatively viewed as a superposition of a thermally driven symmetric circulation with deep water formation at both cold poles and upwelling near the equator, and a haline driven circulation that sees the opposite flow with downwelling near the equator and upwelling at both fresh poles. The haline circulation is much weaker than the thermally driven one as the former is forced by a constant salinity flux while the latter by a restoring boundary condition which imposes a fixed surface temperature. This causes sharper isopycnals and hence a stronger flow \citep{thual1992catastrophe}. Therefore in the pole-to-pole circulation the thermally driven downwelling near one of the poles is stronger than the salinity-induced upwelling near the other. As volume needs to be conserved a small vertical boundary layer for the downwelling is countered by upwelling in the rest of the basin. The horizontal length scale of this layer is roughly $0.9$, which is consistent with the scaling argument presented in \citet{quon1992multiple}.

The stability of these states can be physically explained using arguments presented in \citet{turner1973buoyancy}. In the steady state, diffusion of salt and heat is balanced by buoyancy-induced mixing. If a perturbation at the surface increases the horizontal density gradient, and hence the overturning strength, then more warm and salty water will be transported to the bottom via downwelling, while more cold and fresh water wells up. This water needs to be heated and salinified by the surface forcing via diffusion. If this cannot keep up with the increased supply of deep water, then the horizontal density gradients will fall, and so the overturning strength decreases again. This negative feedback mechanism stabilizes the pole-to-pole circulations.

The saddle state on the other hand is completely thermally driven. Its instability is due to the competition between the two overturning cells: if a perturbation would cause one of the cells to be slightly larger than the other, then the larger cell would see a net input of salt via the freshwater forcing. This would strengthen  its downwelling, and hence this cell would grow even more and eventually become the sole overturning cell. 

\section{Methodology}\label{sec:method}
\subsection{Freidlin-Wentzell Theory}
Consider a SPDE with additive noise
\begin{equation}\label{eq:standardSPDE}
    \partial_t\boldsymbol{U} = \mathcal{B}(\boldsymbol{U}) + \sqrt{\varepsilon}\mathcal{F}(\boldsymbol{\eta})(\boldsymbol{x},t),\qquad t\geq0
\end{equation}
where $\boldsymbol{U}:\,\R^d\times\R_{\geq0}\to\R^m$ is the $m$-dimensional state of the system as a function of space $\boldsymbol{x}\in\R^d$ and time $t\in\R_{\geq0}$. The deterministic drift is given by the operator $\mathcal{B}(\boldsymbol{U})$, and operator $\mathcal{F}(\boldsymbol{\eta})$ acts on the Gaussian spatio-temporal white noise $\boldsymbol{\eta}$, which is scaled by smallness-parameter $\varepsilon$. The covariance operator is denoted by $\mathcal{A} = \mathcal{F}^*\mathcal{F}$, which is the adjoint of $\mathcal{F}$ acting on itself. For simplicity, we only treat additive noise, but the theory can be extended to include multiplicative noise. It is non-trivial to show in general that this SPDE is well-posed, since the spatially rough noise necessitates carefully defining any nonlinearities present in $\mathcal{B}(\boldsymbol{U})$ \citep{hairer2014theory}. Since our choice of spatial noise covariance with a finite number of wave-lengths $K$ introduced a natural cut-off, such complications are avoided, and we can formally extend Freidlin-Wentzell Theory directly to the infinite-dimensional setup \citep{grafke2017long}. Concretely, we are interested in situations where the stochastic process \eqref{eq:standardSPDE} realizes a certain transition within a time $\tau$ starting at $\boldsymbol{U}(\boldsymbol{x},0) = \boldsymbol{U}_0(\boldsymbol{x})$ and ending at $\boldsymbol{U}(\boldsymbol{x},\tau) = \boldsymbol{U}_1(\boldsymbol{x})$. This transition might be impossible in the deterministic setting ($\varepsilon = 0$) but can occur if noise is present ($\varepsilon > 0$) although it might be increasingly rare in the limit of small noise ($\varepsilon\to0$).

LDT allows us to compute the rate at which this probability decays as the size of the noise $\varepsilon$ approaches zero. The probability of observing a realization close to a function $\boldsymbol{\upsilon}:\,\R^d\times[0,\tau]\to\R^m;\,(\boldsymbol{x},t)\mapsto\boldsymbol{\upsilon}(\boldsymbol{x},t)$ obeys
\begin{equation}\label{eq:LDP}
    \mathbb{P}\left[\underset{t\in[0,\tau]}{\text{sup}}\|\boldsymbol{U}(\boldsymbol{x},t) - \boldsymbol{\upsilon}(\boldsymbol{x},t)\|_{L^2}<\delta_R \right] \asymp \exp\left(-S_\tau[\boldsymbol{\upsilon}/\varepsilon]\right)
\end{equation}
for sufficiently small $\delta_R > 0$. Here,  $\asymp$ denotes log-asymptotic equivalence and $\|\cdot\|_{L^2}$ is the $L^2$-norm in the spatial components. The functional $S_\tau[\boldsymbol{\upsilon}]$ is the Freidlin-Wentzell action and is defined as 
\begin{align}\label{eq:FWaction}
    S_\tau[\boldsymbol{\upsilon}] = 
    \begin{cases}
        &\frac{1}{2}\int_0^\tau\left\langle\partial_t\boldsymbol{\upsilon}-\mathcal{B}(\boldsymbol{\upsilon}),\,\mathcal{A}^{-1}\left(\partial_t\boldsymbol{\upsilon}-\mathcal{B}(\boldsymbol{\upsilon})\right)\right\rangle_{L^2}\,dt\qquad\text{if the integral converges}\\
        &\infty\hspace{7.05cm}\text{otherwise}
    \end{cases}
\end{align}
where $\langle\cdot,\cdot\rangle_{L^2}$ denotes the $L^2$ inner product in the spatial components, and $\mathcal{A}^{-1}$ the inverse of the covariance operator. More technical constraints regarding the smoothness of the functions and operators at hand can be found in \citet{freidlin1998random}. In case of degenerate noise,  the covariance operator is not invertible and there are several methods to circumvent the inversion \citep{grafke2019numerical}.

The major implication of \eqref{eq:LDP} is that in the limit of small noise ($\varepsilon\to0$) the trajectory $\boldsymbol{\Tilde{\upsilon}}$ with the smallest action becomes the least unlikely trajectory to realize the transition. In other words
\begin{equation}\label{eq:mini}
    \boldsymbol{\Tilde{\upsilon}}(\boldsymbol{x},t) = \underset{\boldsymbol{\upsilon}(\boldsymbol{x},0) = \boldsymbol{U}_1(\boldsymbol{x}), \,\boldsymbol{\upsilon}(\boldsymbol{x},\tau) = \boldsymbol{U}_2(\boldsymbol{x})}{\argmin}S_\tau[\boldsymbol{\upsilon}]
\end{equation}
is the maximum likelihood pathway (or instanton). Moreover, all realizations that fulfill the transition conditions will concentrate around $\boldsymbol{\Tilde{\upsilon}}$ for low noise levels:
\begin{equation*}
    \underset{\varepsilon\to0}{\lim}~\mathbb{P}\left[\underset{t\in[0,\tau]}{\sup}\|\boldsymbol{U}(\boldsymbol{x},t)-\boldsymbol{\Tilde{\upsilon}}(\boldsymbol{x},t)\|_{L^2}<\delta\,\bigg|\,\boldsymbol{U}(\boldsymbol{x},0) = \boldsymbol{U}_1(\boldsymbol{x}),\,\boldsymbol{U}(\boldsymbol{x},\tau) = \boldsymbol{U}_2(\boldsymbol{x})\right] = 1
\end{equation*}
for $\delta > 0 $ sufficiently small. Our goal now is to compute this instanton $\boldsymbol{\Tilde{\upsilon}}(\boldsymbol{x},t)$ for an overturning collapse in our model. This will be the path whose associated forcing is the most likely to occur out of all possible stochastic forcings that induce a collapse in a time interval $t\in[0,\tau]$.

The minimization \eqref{eq:mini} is a common problem in classical field theory \citep{altland2010condensed}. The Lagrangian of \eqref{eq:FWaction} is 
\begin{equation*}
    \mathcal{L}\left(\boldsymbol{\upsilon},\partial_t\boldsymbol{\upsilon}\right) = \left\langle\partial_t\boldsymbol{\upsilon}-\mathcal{B}(\boldsymbol{\upsilon}),\,\mathcal{A}^{-1}\left(\partial_t\boldsymbol{\upsilon}-\mathcal{B}(\boldsymbol{\upsilon})\right)\right\rangle_{L^2}.
\end{equation*}
Then we can define the conjugate momentum to $\boldsymbol{\upsilon}$ as 
\begin{equation*}
    \boldsymbol{\chi}(\boldsymbol{x},t) = \frac{\partial\mathcal{L}\left(\boldsymbol{\upsilon},\partial_t\boldsymbol{\upsilon}\right)}{\partial\left(\partial_t\boldsymbol{\upsilon}\right)}
\end{equation*}
and so define the Hamiltonian as the Fenchel-Legendre transform of the Lagrangian:
\begin{align*}
    \mathcal{H}\left(\boldsymbol{\upsilon},\boldsymbol{\chi}\right) &= \left(\boldsymbol{\chi}\partial_t\boldsymbol{\upsilon}-\mathcal{L}\left(\boldsymbol{\upsilon},\partial_t\boldsymbol{\upsilon}
    \right)\right)\Big|_{\partial_t\boldsymbol{\upsilon} = \partial_t\boldsymbol{\upsilon}(\boldsymbol{\upsilon},\boldsymbol{\chi})}\\
    &= \langle\mathcal{B}(\boldsymbol{\upsilon}),\boldsymbol{\chi}\rangle_{L^2} + \frac{1}{2}\langle\boldsymbol{\chi},\mathcal{A}(\boldsymbol{\chi})\rangle_{L^2}.
\end{align*}
Hence, the Hamiltonian field equations, also called the instanton equations, follow directly as
\begin{subequations}\label{eq:instantoneq_theory}
\begin{align}
    \partial_t\boldsymbol{\upsilon} &=& \mathcal{B}(\boldsymbol{\upsilon}) + \mathcal{A}(\boldsymbol{\chi})\label{eq:ups}\\
    \partial_t\boldsymbol{\chi} &=& -\left(\nabla_{\boldsymbol{\upsilon}}\mathcal{B}\right)^*(\boldsymbol{\chi})\label{eq:chi}  
\end{align}
\end{subequations}
with boundary conditions $\boldsymbol{\upsilon}(\boldsymbol{x},0) = \boldsymbol{U}_1(\boldsymbol{x})$ and $\boldsymbol{\upsilon}(\boldsymbol{x},\tau) = \boldsymbol{U}_2(\boldsymbol{x})$. Solving these equations yields the instanton $\boldsymbol{\Tilde{\upsilon}}(\boldsymbol{x},t)$. 
The advantage of reformulating the minimization problem into \eqref{eq:instantoneq_theory} is that the covariance operator $\mathcal{A}$ no longer needs to be inverted. A disadvantage of the Hamiltonian framework is that we have one first-order partial differential equation with an initial value as well as a final condition, while the conjugate momentum equation has no temporal boundary conditions. 

\subsection{The instanton equations for the Boussinesq flow}
We formulate the instanton equations \eqref{eq:instantoneq_theory} for the SPDE of the Boussinesq flow \eqref{eq:SPDE}. Let $\bph(x,z,t) = \left(\omega, T, S\right)^T(x,z,t)$ be the instanton path, and $\bt(x,z,t) = \left(\theta_\omega, \theta_T, \theta_S\right)^T(x,z,t)$ its conjugate momentum. For the covariance operator $\mathcal{A}$ we find
\begin{gather*}
    \mathcal{A} = \sigma^*\sigma = 
    \begin{pmatrix}
        0 & 0 & 0 \\
        0 & 0 & 0 \\
        0 & 0 & \frac{(h(z))^2}{\tau_S^2K}\mathcal{U}^*\mathcal{P}\mathcal{U}
    \end{pmatrix}
\end{gather*}
where we used that the adjoint of the basis transformation operator is its inverse ($\mathcal{U}^* = \mathcal{U}^{-1}$), and that $\mathcal{P}$ is a projection operator, so $\mathcal{P}^* = \mathcal{P}$ and $\mathcal{P}^2 = \mathcal{P}$. Note that the covariance operator is clearly non-invertible, as expected: the noise in our model only acts on the salinity component and is therefore degenerate. Essentially $\mathcal{A}(\bt)$ is a filter that yields the Fourier modes of $\theta_S$ that are constant along the vertical axis and have a horizontal wavelength $\Asp/k$ for $1\leq k\leq K$. The amplitude of the noise is indicated by $\varepsilon$ just as in \eqref{eq:standardSPDE}. The set of PDEs describing the instanton $\bph$ of an AMOC collapse in \eqref{eq:SPDE} is
\begin{subequations}\label{eq:phi}
\begin{align}
    &&\partial_t\omega + \partial_z\psi\partial_x\omega-\partial_x\psi\partial_z\omega = \Pran\nabla^2\omega+\Pran\Ray\left(\partial_xT-\partial_xS\right)\\
    &&\partial_tT + \partial_z\psi\partial_xT-\partial_x\psi\partial_zT = \nabla^2T+\frac{h(z)}{\tau_T}\left(T_S(x)-T\right)\\
    &&\partial_tS + \partial_z\psi\partial_xS-\partial_x\psi\partial_zS = \Lew^{-1}\nabla^2S+\frac{h(z)}{\tau_S}S_S(x)\\
    &&\qquad\qquad
    + \frac{(h(z))^2}{\tau_S^2K}\sum_{k=1}^K\theta_k^{(1)}(t)\cos\left(\frac{2\upi}{\Asp}kx\right) + \theta_k^{(2)}(t)\sin\left(\frac{2\upi}{\Asp}kx\right)\nonumber
\end{align}
\end{subequations}
where
\begin{align*}
    &-\nabla^2\psi = \omega\qquad\text{ with }\psi = 0\text{ for }(x,z)\in\partial\Omega\\
    &\theta_k^{(1)}(t) = \frac{2}{\Asp}\int_\Omega\theta_S(x,z,t)\cos\left(\frac{2\upi}{\Asp}kx\right)\,dx\,dz\\
    &\theta_k^{(2)}(t) = \frac{2}{\Asp}\int_\Omega\theta_S(x,z,t)\sin\left(\frac{2\upi}{\Asp}kx\right)\,dx\,dz
\end{align*}
with spatial boundary conditions
\begin{align*}
    &\omega = 0\text{ for }(x,z)\in\partial\Omega\\
    &\partial_xT = \partial_xS = 0\text{ for }x= 0,\,\Asp\\
    &\partial_zT = \partial_zS = 0\text{ for }z= 0,\,1
\end{align*}
and temporal boundary conditions
\begin{align*}
    &\bph(x,z,0) = \bph_\ON\\
    &\bph(x,z,\tau) = \bph_\OFF.
\end{align*}
The set of PDEs for the conjugate momentum $\bt$ is derived by first computing the variational derivatives of the deterministic drift $f$ and their adjoints, see Appendix \ref{appA}. This results in the following equations:
\begin{subequations}\label{eq:theta}
\begin{align}
    &&\partial_t\theta_\omega + \partial_z\psi\partial_x\theta_\omega - \partial_x\psi\partial_z\theta_\omega + \nu_\omega + \nu_T + \nu_S + \Pran\nabla^2\theta_\omega = 0\\
    &&\partial_t\theta_T + \partial_z\psi\partial_x\theta_T - \partial_x\psi\partial_z\theta_T + \nabla^2\theta_T - \frac{h(z)}{\tau_T}\theta_T - \Pran\Ray\partial_x\theta_\omega = 0\\
    &&\partial_t\theta_S + \partial_z\psi\partial_x\theta_S - \partial_x\psi\partial_z\theta_S + \Lew^{-1}\nabla^2\theta_S + \Pran\Ray\partial_x\theta_\omega = 0
\end{align}
\end{subequations}
where the functions $\nu_\omega, \,\nu_T,\,\nu_S:\,\Omega\times\R_{\geq0}\to\R$ obey the Poisson equations
\begin{align*}
    &-\nabla^2\nu_\omega = \partial_x\omega\partial_z\theta_\omega-\partial_z\omega\partial_x\theta_\omega\\
    &-\nabla^2\nu_T = \partial_xT\partial_z\theta_T-\partial_zT\partial_x\theta_T\\
    &-\nabla^2\nu_S = \partial_xS\partial_z\theta_S-\partial_zS\partial_x\theta_S\\
    &\text{where }\nu_\omega = \nu_T = \nu_S = 0\text{ for }(x,z)\in\partial\Omega
\end{align*}
with spatial boundary conditions
\begin{align*}
    &\theta_\omega = 0\text{ for }(x,z)\in\partial\Omega\\
    &\partial_x\theta_T = \partial_x\theta_S = 0\text{ for }x= 0,\,\Asp\\
    &\partial_z\theta_T = \partial_z\theta_S = 0\text{ for }z= 0,\,1.
\end{align*}
Altogether, PDEs \eqref{eq:phi} and \eqref{eq:theta} will yield the most likely path $\bph$ from the ON to OFF state within time $t\in[0,\tau]$ in the low noise limit as $\varepsilon\to0$. The associated forcing that brings this path about is then $\mathcal{A}(\bt)$.

\subsection{Numerical implementation}
As mentioned before, the problem with the previously described PDE's is that equations \eqref{eq:phi} describing $\bph$ have two temporal boundary conditions, while equations \eqref{eq:theta} regarding the conjugate momentum $\bt$ have none. To deal with this we use the Augmented Lagrangian Method in which these equations are reformulated as a control problem \citep{hestenes1969multiplier, schorlepp2022spontaneous}, see Appendix \ref{appB} for more technical details. The PDEs there are solved using finite difference methods. We use a first-order semi-implicit Euler method to integrate the path and the conjugate momentum equations in time. The spatial derivatives are approximated with a second-order central difference approach \citep{veldman1992playing}. We use a non-equidistant grid to discretize $\Omega$ similar as in \citet{dijkstra1995efficient}. There are $M+1$ gridpoints in the horizontal direction and $N+1$ points in the vertical. The gridpoints are defined as 
\begin{align*}
    x_m &= \frac{m}{M}\Asp\hspace{5.95cm}\text{for }m\in\{0,1,\dots,M\}\\
    z_n &= 0.5 + \tanh\left(q\left(\frac{n}{N}-\frac{1}{2}\right)\right)\Big/ \left(2\tanh\left(\frac{q}{2}\right)\right)\qquad\text{for }q=3\text{ and }n\in\{0,1,\dots,N\}.
\end{align*}
This way we have a finer grid near the surface where the forcing is applied. For the spatial resolution we take $(M,N) = (40,80)$ unless otherwise stated. Here priority is given to a high vertical resolution in order to accurately resolve the top boundary layer. Now the bifurcations of the system will shift under different spatial resolutions, but the same qualitative structure will remain \citep{dijkstra1997symmetry}. Similarly, the instanton will show only a quantitative shift under different resolutions but the physical mechanisms will remain the same. Hence, the following analyses can be carried out similarly for any resolution. For the time discretization we employ a step size $\Delta t = 0.01$. 

\begin{figure}
    \includegraphics[scale = 0.25]{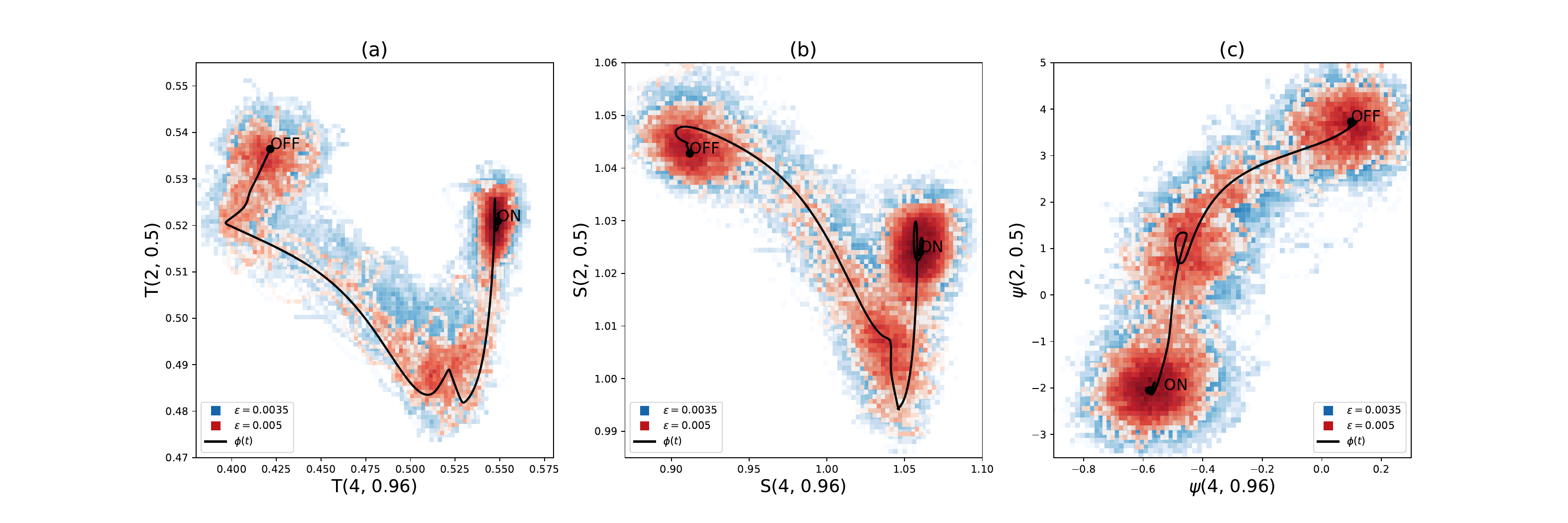}
    \caption{Instanton (black) and histograms of 55 realized transitions (blue) at noise level $\varepsilon = 0.005$ and 10 realized transitions (red) at noise level $\varepsilon = 0.0035$} both with logarithmic scaling in the various phase spaces $\left(T(4,0.96), T(2,0.5)\right)$ (a), $\left(S(4,0.96), S(2,0.5)\right)$ (b) and $\left(\psi(4,0.96), \psi(2,0.5)\right)$ (c) with the ON and OFF state indicated. Model parameters are $\beta = 0.1$ and $(M,N) = (15,30)$.
    \label{fig:verification}
\end{figure}

In order to verify our method, we compare the computed instanton with several realized transitions at low noise levels. As it is computationally demanding to generate these transitions at the standard resolution, we compute the instanton and the realizations at the lower resolution of $(M,N) = (15,30)$. In figure \ref{fig:verification} the resulting instanton is shown in several projected phase spaces together with multiple realized transitions. The instanton lies at the centre of the tube of transitions in salinity and streamfunction spaces, which is a good indication that it is indeed correct and representative of observed stochastic transitions. However, in the temperature variable this agreement is less obvious: transitions projected onto the temperature space are quite erratic compared to those projected onto the salinity and streamfunction spaces. This illustrates that the underlying quasi-potential is relatively flat there, which can cause the most likely transition path for finite noise to deviate from the Freidlin-Wentzell instanton \citep{borner2023saddle}. If we compare the transitions for noise level $\varepsilon = 0.005$ to those with $\varepsilon = 0.0035$, then we see that the latter lie closer and around the instanton. This indicates that for even lower noise the transition tube and instanton also agree in temperature space. Generating unbiased transition paths for even lower noise levels is however unfeasible as for $\varepsilon = 0.0035$ the Monte Carlo estimated probability of tipping within time $\tau = 20$ is roughly $10^{-3}$. All in all, we can justify that the computed path is indeed the Freidlin-Wentzell instanton from the ON to OFF state.

In addition to the method's verification we also need to argue its validity. The instanton will be a fair representation of a noise-induced AMOC collapse if two assumptions hold: (i) the noise is indeed white in time and spatially uniformly represented by the first $K$ Fourier components, and (ii) the system is in the limit of low noise. Assumption (ii) has already been argued for the model used in \citet{soons2023optimal}, and the noise compared to the AMOC strength should not depend too much on the model choice. Moreover, estimates of the annual standard deviation in sea surface salinities (SSS) \citep{friedman2017new} yield $\sqrt{\varepsilon}\in(10^{-3},10^{-2})$. This leads to a variance of $\varepsilon\in(10^{-6},10^{-4})$. As we know from the verification at these noise levels the collapse probability in the low-resolution version of the model is already less than $10^{-3}$. Hence we can safely assume the low-noise limit also holds for this model. The choice of noise structure (i) is harder to justify. For the spatial structure we have taken a straightforward choice of a limited number of sinusoids based on variations in the global mean precipitation patterns. That the noise is Gaussian and white in time is to keep the computations simple, and is standard for other stochastic AMOC models \citep{cessi1994simple, timmermann2000noise, castellana2019transition}. A thorough analysis of the noise structure in the Atlantic Basin can be found in \citet{boot2024}.

\begin{figure}
    \centering
    \includegraphics[width = \textwidth]{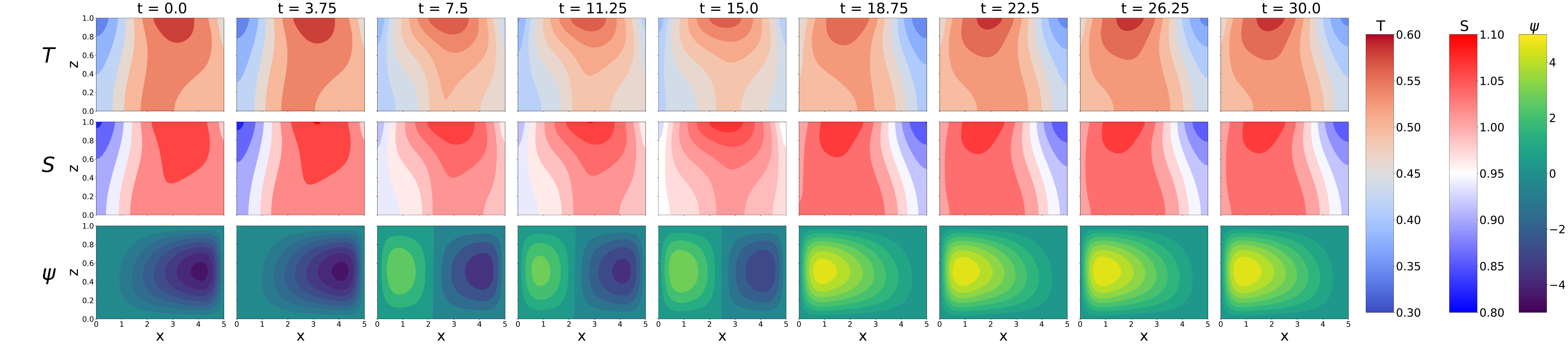}
    \caption{Instanton from $\bph_{\ON}$ to $\bph_{\OFF}$ for $\beta = 0.1$ at times $t\in\{0.0, 3.75, 7.5, 11.25, 15.0,$ $ 18.75, 22.5, 26.25, 30.0\}$ (left to right) with temperature $T$ (top), salinity $S$ (middle) and streamfunction $\psi$ (bottom). }
    \label{fig:overview_inst}
\end{figure}

\section{The most likely overturning circulation collapse}\label{sec:collapse}
\begin{figure}
    \centering
    \includegraphics[scale = 0.3]{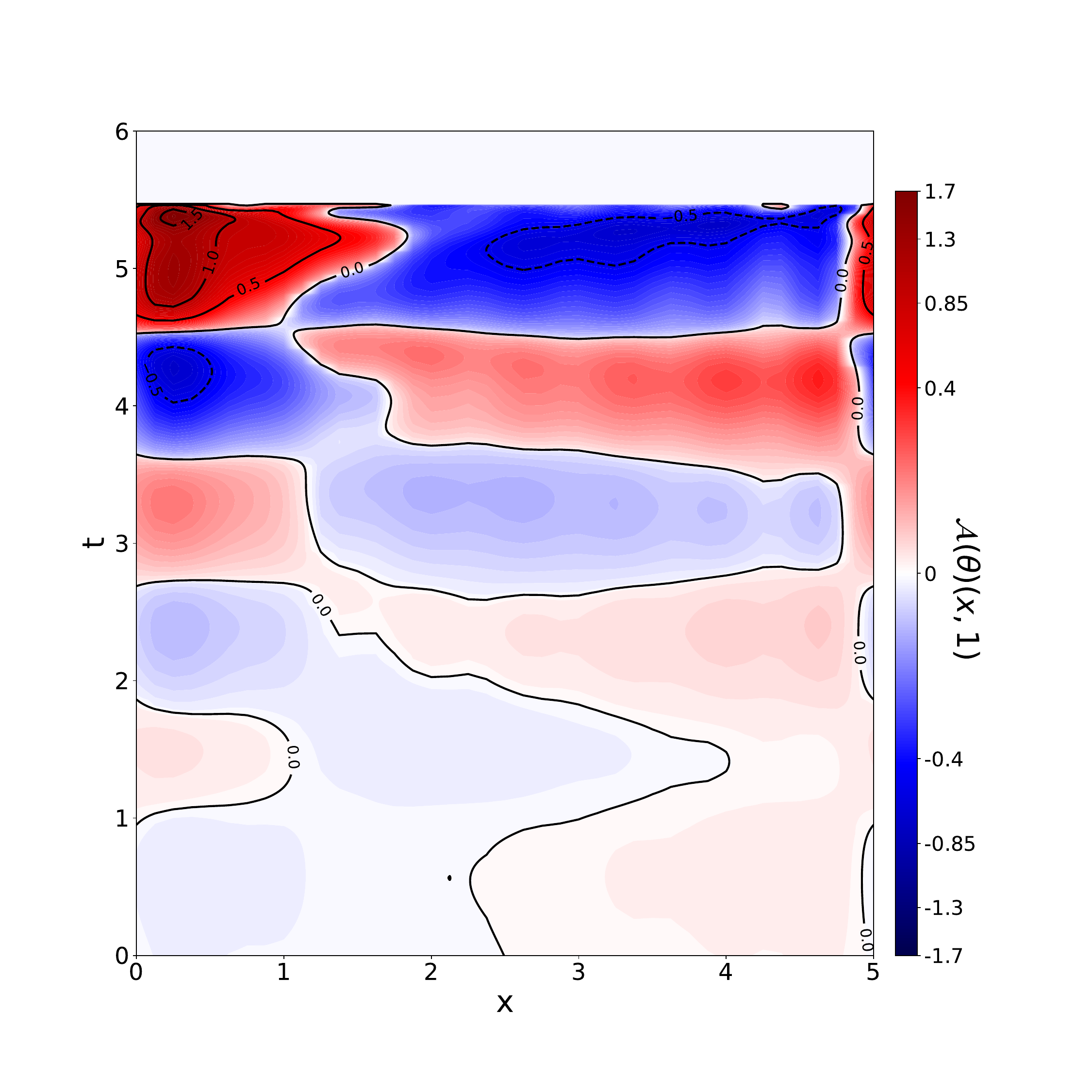}
    \caption{Hovm\"oller diagram of the salinity forcing at the surface ($\mathcal{A}(\bt)(x,1)$) for $t\in[0,6]$. Note that the color mapping is non-linear.}
    \label{fig:overview_fresh}
\end{figure}

In this section we present the most likely transition path from the  ON to the OFF state for $\beta = 0.1$ The value of $\beta$ is chosen close to the bifurcation point $\beta = 0.11$, because a noise-induced transition is more probable close to the bifurcation point and the resulting trajectory therefore more relevant. We allocated a time interval of $\tau = 20$ for the forced part of the trajectory up to the separatrix. For larger time intervals, our procedure yields identical dynamics as result, but with the trajectory initially spending additional time in the ON state. This is evidence that we are already in the long-time limit, and no further effects are expected by allowing a longer transition time. The resulting instanton is shown in figure \ref{fig:overview_inst} together with the associated optimal forcing in the Hovm\"oller diagram in figure \ref{fig:overview_fresh}. Here we have restricted  the instanton to the interval $t\in[0,30]$ where end and beginning are sufficiently close to the OFF and ON state. 

From these figures we get a rough idea of the collapse. The most likely freshwater forcing consists of a repeating pattern of first a freshening in the south ($x\lessapprox 1.5$) followed by a salinification, while in the rest of the basin compensation occurs. The largest forcing peak is reached at $t = 5.38$ and the forcing ceases completely at $t=5.5$ at which point the separatrix is reached. At this point the forcing has created a new additional overturning cell in the south. From thereon the system is in a symmetric two-cell overturning configuration. Note that the temperature and salinity distribution are still skewed towards the north. Only at $t \approx 15$ does the instanton approach the saddle state where both tracers and the streamfunction have a symmetric distribution (the latter is actually anti-symmetric). Subsequently, after passing the saddle, the original northern cell collapses, and the new southern cell grows. Eventually, the system converges to the stable OFF state. We will discuss this process in more detail, where we distinguish three stages: the forcing to create the new cell ($t\in[0,5.5]$), the two-cell configuration leading to the saddle ($t\in[5.5,15]$) and the collapse of the original cell ($t\in[15,30]$). These will be analyzed from a mechanical and an energetics perspective.

\subsection{Fluid dynamical mechanism of the collapse}
\subsubsection{The forcing}
\begin{figure}
    \centering
    \includegraphics[width = \textwidth]{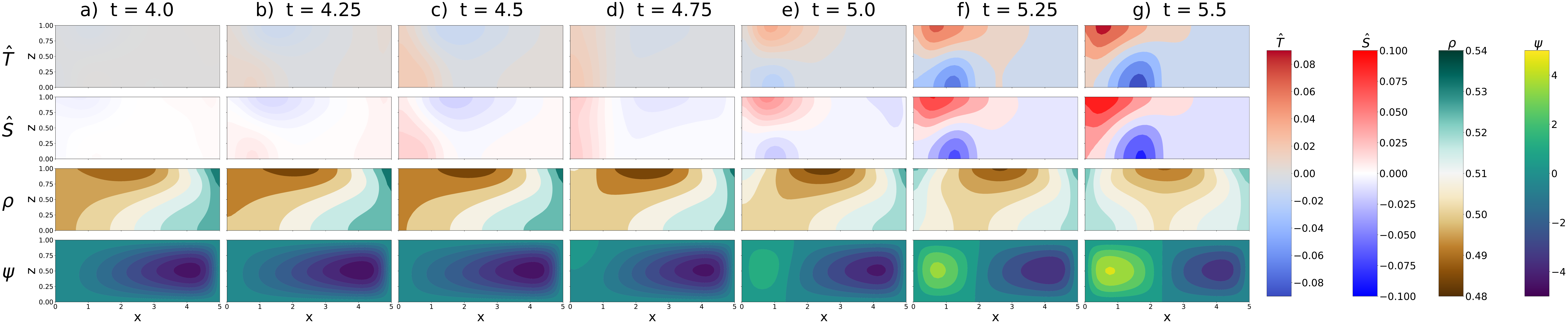}
    \caption{Instanton from $\bph_{\ON}$ to $\bph_{\OFF}$ for $\beta = 0.1$ at times $t\in\{4.0, 4.25, 4.5, 4.75, 5.0, 5.25, 5.5\}$ (left to right) with temperature anomaly $\hat{T} = T - T_{\ON}$ (top), salinity anomaly $\hat{S} = S - S_{\ON}$ (second row), density $\rho$ (third row) and streamfunction $\psi$ (bottom).}
    \label{fig:overview_forcing}
\end{figure}

In figure \ref{fig:overview_fresh} it is shown that the forcing consists of three repeating sequences. We will only discuss the last one
($t\in[4.0,5.5]$), as this one is the largest and in shape similar to the other two. The corresponding instanton trajectory is shown in figure \ref{fig:overview_forcing} where salinity and temperature anomalies ($\hat{S} = S - S_{\ON}$, $\hat{T} = T - T_{\ON}$) and the dimensionless density ($\rho = S-T$) are depicted. Paradoxically the most likely collapse starts off with a strengthening of the pole-to-pole circulation. For $t\in[3.8,4.6]$ a freshwater pulse is added to the southern part of the basin, while the rest of the basin is salinified. This increases the horizontal density gradients and hence the strength of the overturning cell too, which causes that this freshwater anomaly is transported from the south to north, and this salinity anomaly is transported to the bottom, see figure \ref{fig:overview_forcing} b and c. If we take the minimum of $\psi$ as the overturning strength of the northern cell, then the original strength is $\psi_{\text{min},\ON}\approx -4.25$, it peaks at $t = 4.3$ with $\psi_{\text{min}}\approx -4.71$ and returns to its original level at $t=4.74$. However, this latter state is a transient state and its anomalies with respect to the stable ON state are shown in figure \ref{fig:overview_forcing} c and d. It shows that in this state the positive salinity and temperature anomalies have traversed the overturning circulation partially and are now near the bottom of the southern boundary, while their adverse anomalies are near the equator. Looking at the dimensionless density we see that this transient state is less stable: the vertical density gradient in the south has risen to nearly zero.

After the initial southern freshwater pulse and strengthening, a strong salinity forcing is now applied to the southern region for $t\in[4.6,5.5]$, while the rest of the basin is freshened as compensation. One can see in figure \ref{fig:overview_forcing}d that as the initial positive salinity anomaly upwells in the south, the strong salinity perturbation is applied there at the surface. This way a large salinification can be achieved there using multiple smaller perturbations, as we can see from figure \ref{fig:overview_fresh}, which is evidently more likely to occur than one large perturbation. Meanwhile, due to the now weakened upwelling in the south relatively fresh and cool water stays behind near the bottom, causing negative anomalies there (figure \ref{fig:overview_forcing}e). The effects of this large salinification pulse start to appear around $t\approx 5.0$ (figure \ref{fig:overview_forcing}e): a positive temperature and salinity anomaly in the south on top of a negative anomaly. These dipoles induce a new second overturning cell in the south as the heavier water sinks and the lighter rises. Note that this temperature dipole pattern has a dampening effect on the new cell while the salinity dipole aides the new cell. At $t = 5.5$ (figure \ref{fig:overview_forcing}g) the new cell is strong enough to maintain its own density gradients, and the two-cell configuration is established. The forcing is no longer needed and from hereon the system will reach the OFF state in a deterministic way.

\begin{figure}
    \centering
    \includegraphics[width = \textwidth]{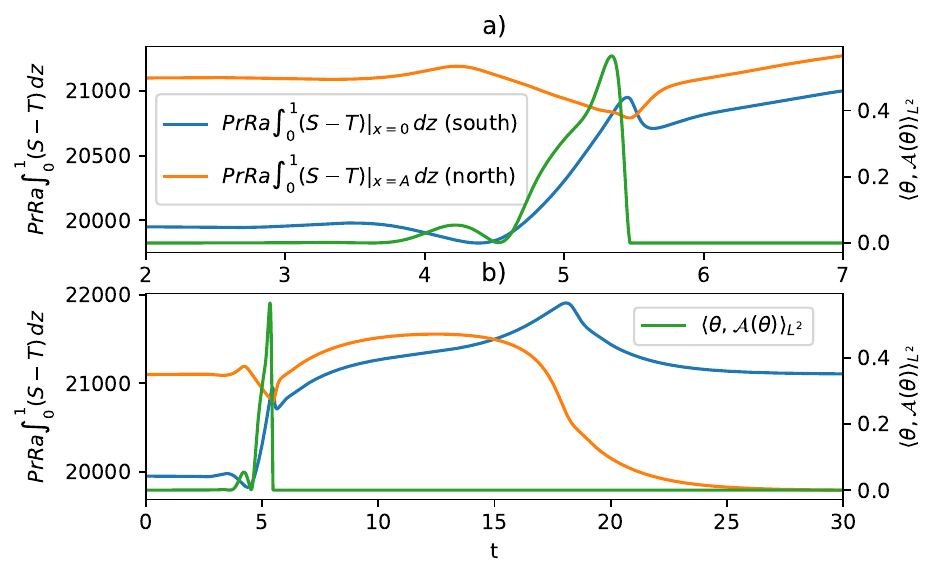}
    \caption{The cumulative density $\Pran\,\Ray\int_0^1\rho\,dz$ in the south ($x=0$, blue) and the north ($x=\Asp$, orange) (left axis) and the total forcing ($\langle\theta,\mathcal{A}(\theta)\rangle_{L^2}$, green) (right axis) for $t \in(2,7)$ (top) and $t\in(0,30)$ (bottom).}
    \label{fig:cumdensity_forcing}
\end{figure}

Regarding the applied optimal forcing we will discuss three aspects: its magnitude, its timing, and its location. As the instanton is the path that is the most efficient to reach the OFF state in terms of applied forcing, we expect that the created density anomalies to be just large enough to generate the second cell. We consider the volume integrated vorticity equation \eqref{eq:momentum}
\begin{align*}
    \der{}{t}\int_\Omega \omega\,dx\,dz =& \Pran\left[\int_0^1\partial_x\omega\big|_{x=\Asp}- \partial_x\omega\big|_{x=0}\,dz 
    + \int_0^\Asp\partial_z\omega\big|_{z=1}-\partial_z\omega\big|_{z=0}\,dx\right]\\
    &+ \Pran\Ray\int_0^1 (S-T)\big|_{x=0} - (S-T)\big|_{x=\Asp}\,dz
\end{align*}
with the two terms on the right-hand side representing diffusion and buoyancy forcing respectively. To reach a two-cell configuration we need a second cell that is roughly as strong as the first one i.e. the negative vorticity blob in the north has to be countered by a similar but positive vorticity blob in the south. Due to this symmetry, the terms representing the diffusion cancel out and the volume-integrated vorticity balance for the quasi-stable two-cell configuration becomes
\begin{align*}
    0 \approx \Pran\Ray\int_0^1\rho\big|_{x=0} - \rho\big|_{x=\Asp}\,dz
\end{align*}
with dimensionless density $\rho = S-T$. So the required density anomalies are such that the cumulative density at the southern boundary is as large as the one at the northern boundary. Intuitively one can argue that equal densities at the meridional boundaries imply downwelling of equal strength near the poles and hence two overturning cells of equal strength. In figure \ref{fig:cumdensity_forcing} these depth-integrated densities near the poles are shown together with the total forcing $\langle\bt,\mathcal{A}(\bt)\rangle_{L^2}$. It can be seen that as soon as the southern depth-integrated density is as large as the northern one (at $t = 5.35$) the forcing rapidly decreases to zero: the newly formed cell is strong enough and the forcing is no longer needed. Note that, as the forcing decreases, the southern density overshoots the northern density, and at $t = 5.5$ --as the forcing has ceased-- the cumulative densities are equal again. From thereon, as the instanton continues deterministically, the southern density is again lower than the northern density, but it is now significantly higher than originally in the ON state, as there is now downwelling near the southern pole. This continues until they are again equal when the saddle state is reached at $t\approx 15$, after which the northern cell collapses. 

Note that beforehand we see the previous initial strengthening in action with the cumulative density in the south reaching a local minimum around $t\approx 4.3$. For $t\in[5.5,15]$ we have $\int_0^1\rho_{x=0}\,dz < \int_0^1\rho_{x=\Asp}\,dz$, which seems to partially contradict the previous reasoning, but the cause is that the two-cell configuration is still slightly asymmetric in the vorticity 
and so the assumptions do not completely hold. During this time interval the new cell is slightly smaller than the original one but 
it can still sustain itself after its establishment.   

Considering the timing of the forcing it is found that the alternating pulses each last for roughly $0.90\pm0.05$. As discussed, a positive (negative) salinity anomaly created by a pulse in the southern part of the basin weakens (strengthens) the original cell, resulting in an decreased (increased) salinity concentration at the bottom of the southern part. This upwells as a new alternate pulse is applied to the southern surface. In this way two alternating pulses constitute the effect of one big perturbation. As seen in figure\ref{fig:overview_forcing}b-e the upwelling of the anomaly in the south in the ON state takes roughly $0.9$ in time, which 
explains the observed frequency. 

Lastly, regarding the location of the forcing we see that the forcing alternates sign around $x \approx 1.5$ with the eventual goal of creating a large salinity concentration in this part of the domain (figure \ref{fig:overview_forcing}). Now, in order to create a second overturning cell in the south we need a positive vorticity blob which is induced by strong negative salinity gradients. For $\beta = 0.11$ the deterministic salinity forcing gradient $\partial_xS_S$ has its minimum for $x\approx 1.3$. The optimal stochastic salinity forcing has its largest gradient also around this point as it switches sign, exacerbating the negative salinity gradient and hence increasing the vorticity in the southern part of the basin.

\subsubsection{The two-cell configuration}
\begin{figure}
    \centering
    \includegraphics[width = \textwidth]{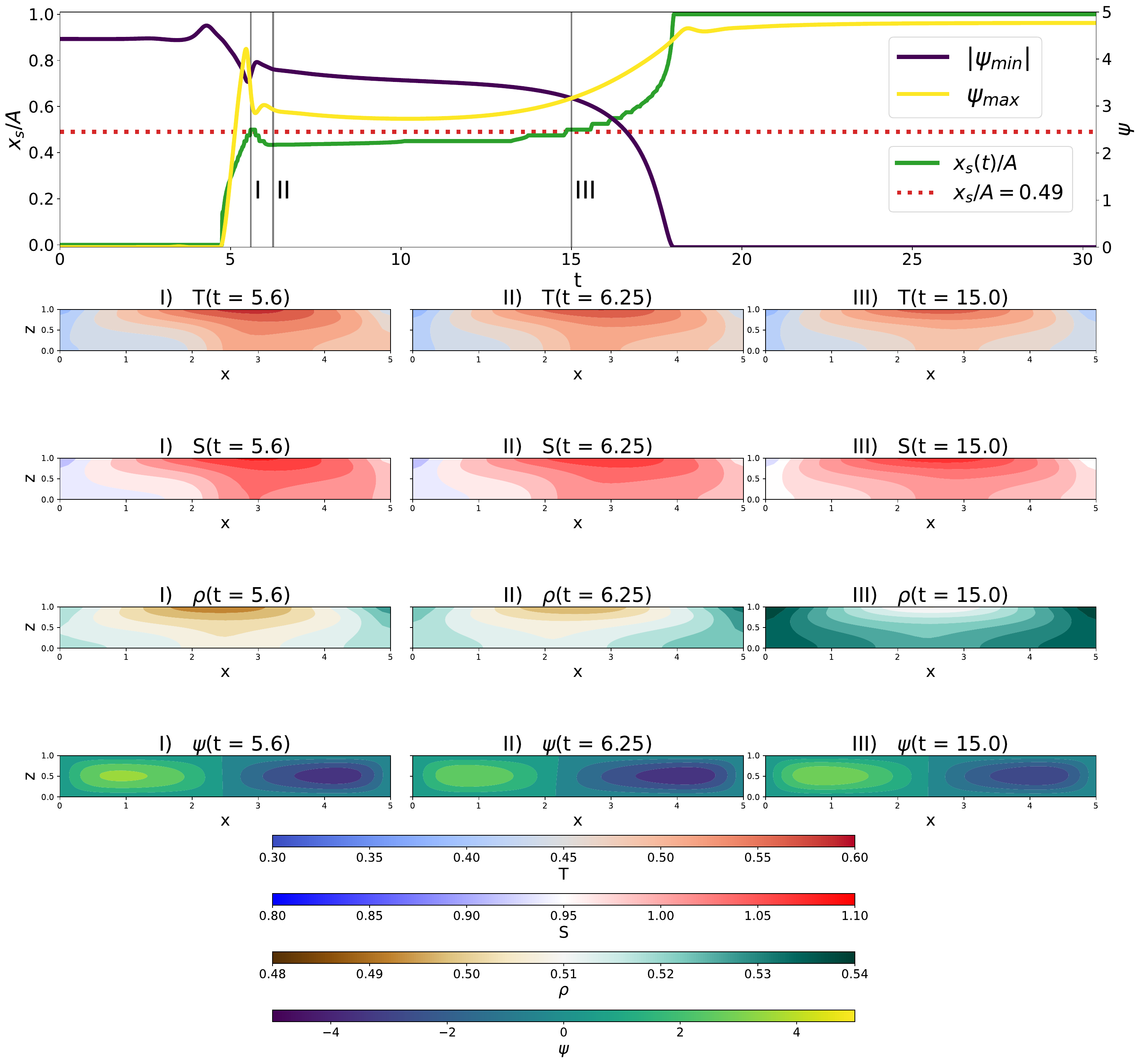}
    \caption{Top: $x_s(t)/\Asp$ (green) along the instanton together with saddle state values $x_s/\Asp = 0.49$ (red, dashed) on the left axis, and $\|\psi_{\text{min}}\|$ (purple) and $\psi_{\text{max}}$ (yellow) indicating the strength of the northern and southern cell respectively on the right axis. States $I$ and $II$ are indicated and visualized in the left and right column respectively with temperature $T$ (top row), salinity $S$ (second row), density $\rho$ (third row) and streamfunction $\psi$ (bottom row).}
    \label{fig:2cell_overview}
\end{figure}

From $t=5.5$ to $t=15.5$ the instanton is in a quasi-stable two-cell configuration. An overview is provided in figure \ref{fig:2cell_overview}, where $\psi_{\text{min}}$ and $\psi_{\text{max}}$ indicate the strength of the northern and southern overturning cell respectively, and $x_s$ is the vertically averaged horizontal coordinate away from the boundaries where $\psi\approx0$ i.e. it is the approximate boundary between the two overturning cells. For the saddle state it can be deduced that $x_s \approx 0.49\Asp$ for $\beta = 0.1$ since at this value the deterministic salinity forcing to each cell is equal. 

From figure \ref{fig:2cell_overview} it follows that initially after the stochastic forcing has ceased the new southern cell reaches a local maximum in size and strength before weakening, see state \rom{1} at $t = 5.6$. To start the new cell the density near the southern surface has been increased by the salinity forcing. As this sinks, its strength peaks followed by a peak in cell size. Now this surface water is not directly replaced by new equally dense water, as can be seen in the density distribution of state \rom{1}, where dense water has sunk to the bottom in the south without dense water following. This causes the intermediate weakening. Once at $t\approx6$ the densities in the southern cell have redistributed into this quasi-stable state \rom{2}. 

Interestingly at state \rom{2} the southern cell is both smaller and weaker than the northern cell, but nevertheless grows, so that eventually the instanton reaches the symmetric saddle state \rom{3} at $t = 15$. Note that in state \rom{2} the salinity and temperature distribution are still skewed towards the north, but they complement each other such that the density distribution and consequently the streamfunction are roughly symmetric around the equator. Moreover, the overall density is lower than in the saddle state \rom{3}. To examine how the weaker and smaller southern cell can grow, we formulate the various transports of density into both cells. With $\Omega_S = [0,x_s]\times[0,1]$ and $\Omega_N = \Omega-\Omega_S$ we denote the southern and northern cell, respectively, and $F^i$ and $Q^i$ stand for the salinity and heat transport into cell $i$. For the salinity transports we have
\begin{align*}
    \text{deterministic surface salinity transport}\quad\quad F_d^S &= \frac{1}{\tau_S}\int_{\Omega_S}h(z)S_S(x)\,dx\,dz\\
    &\approx \frac{\delta}{\tau_S}\int_0^{x_s}S_S(x)\,dx\\
    \text{advective salinity transport}\quad\quad F_a^S &= \int_0^1\partial_z\psi S\big|_{x=x_s}\,dz\\
    \text{diffusive salinity transport}\quad\quad F_f^S &= \int_0^1\partial_xS\big|_{x=x_s}\,dz\\
    \text{stochastic salinity transport}\quad\quad F_s^S &= \int_{\Omega_S}\mathcal{A}(\theta)\,dx\,dz
\end{align*}
where it holds that $F^S_j = -F^N_j$ due to conservation of salinity, where $j\in\{d,a,f,s\}$ indicates the transport process. Moreover, as $\psi(x=x_s)\approx 0$, we have that the advective transport is negligible. For the heat transports $Q$ we have similar formulations. Note that here we also can neglect advective transport and that $Q_f^S = -Q_f^N$, but that $Q_d^S$ and $Q_d^N$ do not necessarily sum to zero due to the nature of the restoring boundary condition. The total salinity and heat transport into $\Omega_N$ are then $F^N = -F^S_d -F^S_f - F^S_s$ and $Q^N = Q^N_d -Q^S_f$. 

\begin{figure}
    \centering
    \includegraphics[width = \textwidth]{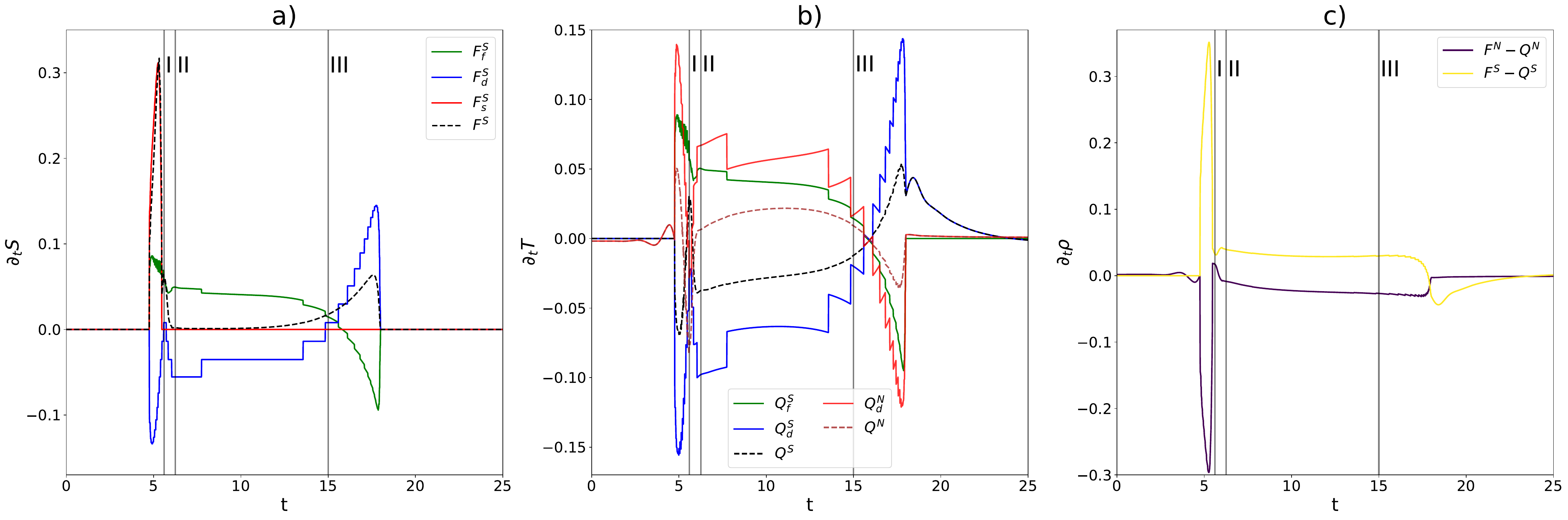}
    \caption{a) The salinity transports into $\Omega_S$ with diffusive transport $F_f^S$ (green), deterministic transport $F_d^S$ (blue), stochastic transport $F_s^S$ (red) and the total $F^S$ (black, dashed). b) The heat transports into $\Omega_S$ with diffusive transport $Q_f^S$ (green), deterministic transport $Q_d^S$ and total $Q^S$ (black, dashed), and into $\Omega_N$ with deterministic transport $Q_d^N$ (red) and total $Q^N$ (brown, dashed). c) The total transport of density into the northern cell $F^N-Q^N$ (yellow) and into the southern cell $F^S-Q^S$ (purple). States \rom{1}, \rom{2} and \rom{3} as in figure \ref{fig:2cell_overview} are indicated.}
    \label{fig:2cell_transports}
\end{figure}

The results are shown in figure \ref{fig:2cell_transports} which illustrates  that the salinity dynamics are responsible for reaching the two-cell configuration of state \rom{1} while the temperature dynamics are the most significant in reaching the saddle state \rom{3}. Examining figure \ref{fig:2cell_transports}a shows that during the two-cell interval ($t\in[5.5,15]$) there is diffusive transport of salt from north to south (as most of the salt is still left in the northern part of the basin), which is almost completely compensated by the deterministic salinity flux at the surface (as $x_s<0.49\Asp$). As the cell grows, the deterministic salt flux at the surface and the diffusive exchange approach zero and then switch sign, where the former slightly outpaces the latter, so that eventually there is a net salt influx from the north to the south. However, the almost zero net salinity exchange ($F^S\approx10^{-4}$) between the two cells during the bigger part of the interval cannot cause the growth of the southern overturning cell. The temperature dynamics in figure \ref{fig:2cell_transports}b are therefore more interesting, where we can see that from \rom{2} to \rom{3} that there is a net southward diffusive transport of heat as most of it is still in the north as a remnant of the original ON state, similar to salinity. However, the southern cell sees a cooling by the surface forcing while the northern cell heats up, which for both cells outweighs the diffusive transport. This is a result of the northern cell's surface being exposed to the hot equator with only the north pole cooling it, whereas the southern cell is only exposed to the cold south pole. As the latter cell grows, we see that both $Q_d^S$ and $Q_d^N$ approach zero and then switch signs. 

The net effect of both heat and salinity transports is that already starting at state \rom{2} there is a net density input into the southern cell and a slight density transport  out off the northern cell. The smaller southern cell has a  disadvantage with regard to the freshwater surface forcing, but this is completely compensated by the diffusive southward salinity transport. The surface heating on the other hand will increase the smaller cell's density and decrease it of the larger cell.  This forcing cannot be compensated by diffusive transport. The key difference is that the salinity forcing is a constant flux applied to the surface, whereas the surface heating is a restoring force. All in all, the new weaker cell sees a net mass import and hence it can grow in size and strength. Eventually the cumulative densities at each pole grow to be equal, see figure \ref{fig:cumdensity_forcing} and the saddle state is reached.

Interestingly, note that at \rom{2} the most destabilizing tracer  is temperature. The streamfunction is close to the symmetric two-cell configuration, and the salt disparity between the two cells is held up by the constant surface forcing. Only the restoring temperature forcing nudges the state towards the saddle. Moreover, as we approach the saddle there has been an overall mass increase. Due to the two-cell state less flow has passed through the equator, and hence the heating has been less effective, and so the whole basin in total experiences a net cooling. Indeed, as seen in figure \ref{fig:2cell_overview},  the density distribution in \rom{3} is still symmetric but elevated.

\subsubsection{The collapse of the northern cell}
At $t \approx  15$ the instanton passes near the saddle state. Beforehand this cannot necessarily be expected: as discussed in \citet{soons2023optimal} the finite-time instanton might be a better representation of the transition path in the low noise limit than the infinite-time instanton. This latter one will pass through the saddle, but --in case the transitions avoid the saddle \citep{borner2023saddle}-- is not necessarily a good exemplar. For this model the transitions nicely follow the instanton as it passes near the saddle, which can be seen in figure \ref{fig:verification} where the three distinct equilibria are the regions where the realizations linger. Therefore saddle-avoidance is no issue here, and the saddle state indeed mediates the transition.

Around $t\approx18$ the original northern cell collapses as $x_s\to\Asp$. In figures \ref{fig:2cell_transports} a and b we can see that leading up to this collapse now both heat and salt transport play a role. As the southern cell expands beyond the equator, the surface salinity forcing aides it, while the surface heating inhibits it. Both are again countered by diffusive transports, but now the diffusive salt transport cannot keep up with the surface forcing as the salinity distribution is no longer skewed but symmetric unlike earlier (during $t\in[5.5,15]$). So as the southern cell grows, more salt and heat is transported southward and their distributions get more skewed, as visible in figure \ref{fig:overview_inst}. The salinity contribution is larger ($F^S>|Q^S|$) and so the southern surface densifies, which aids downwelling there and hence increases the southward transport, which is a signature of  the salt-advection feedback. Eventually the southern cell has grown to fill the whole basin and the OFF-state is almost reached with upwelling in the north where the water is cold and fresh and downwelling in the south where the water is warm and saline.

From $t \approx 18$ the pole-to-pole circulation converges to the OFF-state by a net heating ($Q_S>0$). During the two-cell configuration less water passed through the equator which provided a net cooling of the whole basin. Now that there is again a cross-equatorial flow a net heating occurs, and the overall density of the system decreases (see figure \ref{fig:2cell_transports}c). The overall strength of the OFF-state is slightly larger than the ON-state's ($\psi_{\text{max}}\big|_{t=30}\approx4.77$ versus $\left|\psi_{\text{min}}\right|\big|_{t=0}\approx4.42$) as the salinity forcing is skewed towards the south. 

Lastly, note that after the collapse of the northern cell there is a slight overshoot in overturning strength, with a local maximum $\psi_{\text{max}}\big|_{t=18.4}\approx4.65$ which drops to $\psi_{\text{max}}\big|_{t=18.9}\approx4.59$ before reaching the OFF-state. This decrease is caused by the relatively flat isopycnals in the north just after the collapse which inhibit upwelling there. Once these anomalies are mixed out, and a net heating has taken place the OFF-state is attained.

\subsection{Energetics of the collapse}
The instanton trajectory is also analyzed in terms of its energetics. We use the framework devised by \citet{winters1995available} for density-stratified Boussinesq fluid flows. Similar approaches have been applied to the meridional overturning circulation \citep{hughes2009available, hogg2013energetics}. The non-dimensional volume-integrated kinetic and potential energy are
\begin{align*}
    E_k &= \frac{1}{2}\int_\Omega u^2+w^2\,dx\,dz = \frac{1}{2}\int_\Omega \psi\omega\,dx\,dz\\
    E_p &= \Pran\Ray\int_\Omega\rho z\,dx\,dz,
\end{align*}
where the kinetic energy has been rewritten to avoid derivatives of the streamfunction so numerical discretization errors are minimized. We also use the background potential energy $E_b$ i.e. the minimum potential energy attainable through an adiabatic redistribution of the density. Let $z_*(x,z,t)$ indicate the vertical position in this reference state of the fluid (the background state). The available potential energy $E_a$ is the potential energy released in an adiabatic transition to this background state. So
\begin{align*}
    E_b &= \Pran\Ray\int_\Omega\rho z_*\,dx\,dz\\
    E_a &= E_p-E_b = \Pran\Ray\int_\Omega\rho(z-z_*)\,dx\,dz.
\end{align*}
This reference position $z_*(x,z,t)$ --that is the vertical position of the infinitesimal element at $(x,z,t)$ with density $\rho(x,z,t)$ after an adiabatic reshuffle to the background state-- is computed as 
\begin{align*}
    z_*(x,z,t) = \frac{1}{\Asp}\int_\Omega \mathcal{H}\left(\rho(x',z',t)-\rho(x,z,t)\right)\,dx'\,dz'
\end{align*}
with the Heaviside step function
\begin{align*}
    \mathcal{H}\left(\rho'-\rho\right) = 
    \begin{cases}
        &0\qquad \rho'<\rho\\
        &\frac{1}{2}\qquad\rho'=\rho\\
        &1\qquad\rho'>\rho.
    \end{cases}
\end{align*}

The energy balances are then found by applying the same method as \citet{winters1995available} to our 
model equations. The kinetic energy obeys
\begin{align*}
    \der{}{t}E_k &= \Phi_z - \mathcal{D}\\
    &\Phi_z = \Pran\Ray\int_\Omega\psi\partial_x\rho\,dx\,dz\\
    &\mathcal{D} = \Pran\int_\Omega\omega^2\,dx\,dz
\end{align*}
where $\Phi_z$ denotes the buoyancy flux and $\mathcal{D}\geq0$ the dissipation. For the potential energy we have
\begin{align*}
    \der{}{t}E_p &= \Phi_i - \Phi_z - Q_T\\
    &\Phi_i = -\Pran\Ray\left(\frac{1}{\Lew}\int_0^\Asp S(x,1) - S(x,0)\,dx - \int_0^\Asp T(x,1) - T(x,0)\,dx\right)\\
    &Q_T = \frac{1}{\tau_T}\int_\Omega zh(z)\left(T_S(x)-T\right)\,dx\,dz\\
    &\hspace{0.5cm} \approx \frac{\delta}{\tau_T}\int_0^\Asp T_S(x) - T(x,z)\,dx
\end{align*}
where $\Phi_i$ is the conversion rate of internal energy to potential energy, and $Q_T$ is the net heating by the surface-forcing. Then for the background potential energy 
\begin{align*}
    \der{}{t}E_b &= \Phi_a + \Phi_d\\
    &\Phi_a = \Pran\Ray\int_\Omega z_*\left(\frac{h(z)}{\tau_S}S_S(x)+\mathcal{A}(\bt)-\frac{h(z)}{\tau_T}(T_S(x)-T)\right)\,dx\,dz\\
    &\hspace{0.5cm}\approx\Pran\Ray\Big(\frac{\delta}{\tau_S}\int_0^\Asp z_*(x,1)S_S(x)\,dx - \frac{\delta}{\tau_T}\int_0^\Asp z_*(x,1)(T_S(x)-T(x,1))\,dx\\
    &\hspace{0.7cm}+\int_\Omega z_*\mathcal{A}(\bt)\,dx\,dz\Big)\\
    &\Phi_d = \Pran\Ray\int_\Omega \partial_Tz_*\|\nabla T\|^2 - \frac{1}{\Lew}\partial_Sz_*\|\nabla S\|^2\,dx\,dz 
\end{align*}
where $\Phi_a$ is the rate of change of $E_b$ due to addition of salinity or heat near the surface, while $\Phi_d$ denotes its increase due to diapycnal mixing. Note that $\partial_Tz_* >0$ and $\partial_Sz_*<0$ as $z_*$ is the stably stratified reference state, and hence $\Phi_d\geq0$ means that due to mixing the background potential energy always increases. For the available potential energy budget we simply subtract the background budget from the potential energy budget yielding
\begin{align*}
    \der{}{t}E_a = \Phi_i -\Phi_z -\Phi_a - \Phi_d - Q_T.
\end{align*}
All fluxes and energies are directly computed apart from $\Phi_d$, which we compute as a residual from the background potential energy budget.

\begin{figure}
    \centering
    \includegraphics[width=\textwidth]{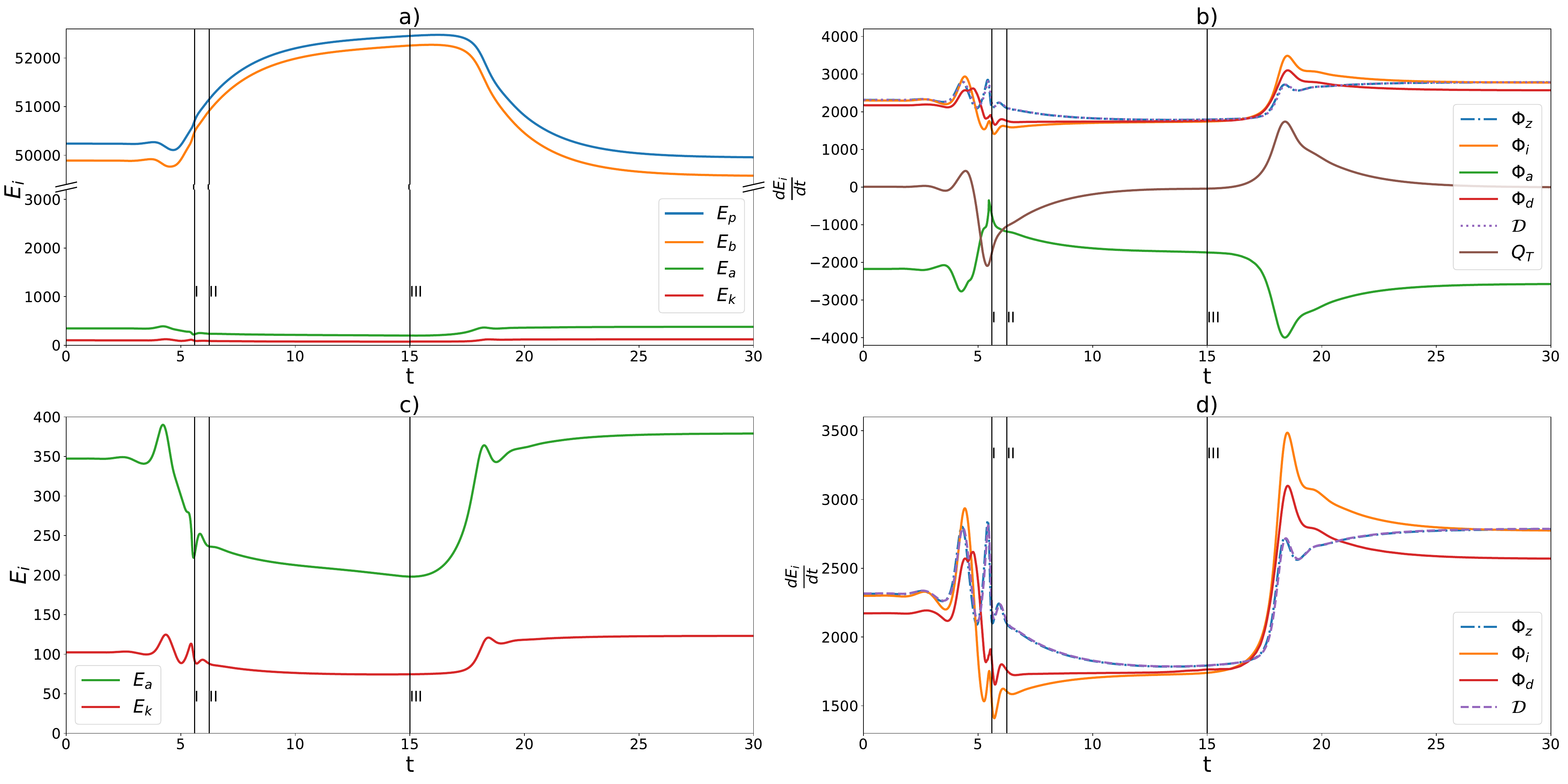}
    \caption{The energies $E_p$ (blue), $E_b$ (yellow), $E_a$ (green) and $E_k$ (red) (a), with zoomed in portion with $E_a$ and $E_k$ (c), and the fluxes $\Phi_z$ (blue, dashed), $\Phi_i$ (yellow), $\Phi_a$ (green), $\Phi_d$ (red), $\mathcal{D}$ (purple, dashed) and $Q_T$ (brown) (b) with zoomed in portion with $\Phi_z$, $\Phi_i$, $\Phi_d$ and $\mathcal{D}$ (d). States \rom{1}, \rom{2} and \rom{3} as in figure \ref{fig:2cell_overview} are indicated.}
    \label{fig:energies}
\end{figure}

An overview of the energies and fluxes along the transition is shown in figure \ref{fig:energies}. The kinetic energy is solely driven by the buoyancy forcing ($\Phi_z$) which is followed by a slightly delayed dissipation ($\mathcal{D}$). Initially it peaks at $t=4.3$ as the original pole-to-pole circulation is strengthened, followed by another peak at $t=5.5$ as the second cell has just formed. Both peaks coincide with the peaks in the forcings $\Phi_z$ and $\mathcal{A}(\bt)$ (figure \ref{fig:cumdensity_forcing}). As these cease,  $E_k$ falls and state \rom{1} is reached. A slight adjustment follows to state \rom{2} and from hereon both the buoyancy forcing and dissipation decrease slowly as the tracers become more evenly distributed. The kinetic energy stays level, and is lower in the two-cell configuration than in the pole-to-pole setting as flow speeds are generally reduced. The kinetic energy budget does not change as the saddle is passed. When the northern cell collapses the buoyancy forcing rises and overshoots since now the circulation is again thermally and haline driven. With it the dissipation and kinetic energy also follow. 

The evolution of the potential energy $E_p$ is relatively simple. At the beginning ($t = 4.7$) it attains a minimum during the transient state as the vertical density gradients in the south approach zero. From thereon it rises steadily until the collapse of the northern cell at $t \approx 18$. Looking at the fluxes we can conclude that the initial minimum is caused by a rise in heating $Q_T$, while the increasing buoyancy forcing and internal mixing $\Phi_i$ roughly cancel each other out. This rise in heating is caused by the surface water being cooler than in the ON-state (figure \ref{fig:overview_forcing}d,  around $t=4.7$). From thereon --because there is no cross-equatorial flow-- a net cooling occurs, and so the overall basin mass and $E_p$ increase. The slightly lower $\Phi_z$ and $\Phi_i$ play a lesser role in this rise. The internal mixing $\Phi_i$ is lowered as diffusion is not as large when the basin contains  two overturning cells. Around $t=18$ this build up reservoir of potential energy is released in the collapse. Via the risen $\Phi_i$ part of this reservoir is converted into internal energy \citep{batchelor2000introduction}, while the increased $\Phi_z$ converts another part into kinetic energy, and a final part is lost to the atmosphere via surface heating $Q_T > 0$.

A more detailed view of the energy transfer is obtained by  splitting the potential energy into available and background potential energy. The background potential energy has roughly the same evolution as the potential energy, meaning that indeed most of the $E_p$-variation is caused by changes in the basin's mass. This is turn is caused by changes in the energy flux from the surface $\Phi_a$, which includes the atmospheric heating. And for the diffusive flux $\Phi_d$ we see similar behaviour as for the internal mixing $\Phi_i$: both peak during the forcing stage and are lowered during the two-cell stage as diffusion is less prevalent. The build-up of $E_b$ is released during the collapse via the atmosphere and diffusive mixing $\Phi_a+\Phi_d$ as available potential energy. The available potential energy $E_a$ shows a peak when $E_b$ has a minimum at $t=4.7$ as the vertical density gradients in the south are small i.e. there is relatively dense water near the surface. This build-up of $E_a$ is released when the second overturning cell is formed and state \rom{1} is reached. During the two-cell stage $E_a$ slowly decreases as the density distribution becomes more symmetric and the isopycnals sharper. At the collapse and the emergence of the pole-to-pole circulation we see dense water moving across the equator and $E_a$ to increase.

In summary the main energetics during the equilibria are an energy input from the atmosphere as available potential energy via surface flux $\Phi_a$ and via the same flux removal of background potential energy to the atmosphere. Then via diffusive fluxes $\Phi_i$ and $\Phi_d$ available potential  energy is again removed as either internal energy or background potential  energy. A last part is removed as buoyancy forcing $\Phi_z$ which then finally sustains the actual circulation and provides kinetic energy. This last one is then again dissipated away as internal energy. Now the forcing puts this system in an imbalance as there is now additional input from the surface via $\Phi_a$ and heating $Q_T$ which cannot be removed fast enough via either diffusive processes or buoyancy fluxes. Eventually, right before the collapse, an unstable balance is reached where the surface input has decreased to a level that can be upheld again by diffusion and buoyancy fluxes. However, a small perturbation to this two-cell configuration increases diffusion, which in turn releases the build-up potential energy, which is converted into kinetic energy and eventually causes the collapse.

\begin{figure}
    \centering
    \includegraphics[scale = 0.13]{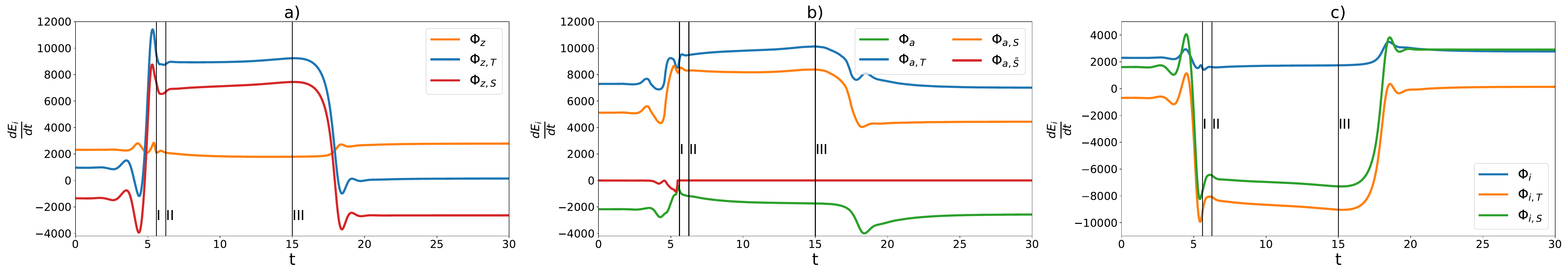}
    \caption{Fluxes $\Phi_z$ (a), $\Phi_a$ (b) and $\Phi_i$ (c) split into a salinity and temperature contribution with the former split again into a stochastic and deterministic component. States \rom{1}, \rom{2} and \rom{3} as in figure \ref{fig:2cell_overview} are indicated.}
    \label{fig:fluxsplit}
\end{figure}

The effects of both tracers on the energetics of the collapse can be determined by splitting  the fluxes $\Phi_z$, $\Phi_i$ and $\Phi_a$  into contributions due to salinity $S$ and temperature $T$, where $\Phi_a$ also has a deterministic and stochastic salinity component $\Phi_{a,S}$ and $\Phi_{a,\Tilde{S}}$. We denote them as
\begin{align*}
    \Phi_z &= \Phi_{z,T} - \Phi_{z,S}\\
    \Phi_a &= \Phi_{a,S} - \Phi_{a,T} + \Phi_{a,\Tilde{S}}\\
    \Phi_i &= \Phi_{i,T} - \Phi_{i,S}
\end{align*}
and their dynamics are shown in figure \ref{fig:fluxsplit}. During the pole-to-pole circulations we have $\Phi_{z,T} >0$ and $\Phi_{z,S}<0$ confirming that these circulations are driven by both salinity and temperature. Salinity aides the upwelling near one pole, while temperature aides the downwelling near the other. During the two-cell configuration both cells are only thermally driven, and indeed the salinity component $\Phi_{z,S}$ has switched sign opposing the creation of kinetic energy. For the components of $\Phi_i$ we observe similar behaviour. The diffusion of salt (heat) aides the transfer of internal energy to potential energy when $\Phi_{i,S}>0$ ($\Phi_{i,T}<0$) and vice versa. So during the pole-to-pole circulations both components help while for two cells only heat diffusion aides. During the two-cell stage the surface is (spatially averaged) much saltier than the bottom as less salt is transported downward by the cells from the equator. Therefore vertical diffusion of salt would reduce the potential energy of the system, and hence less energy would be available for the driving buoyancy flux. On the other hand the surface is much cooler, so vertical diffusion of heat would still aide the potential energy. For the pole-to-pole case for both salt and heat diffusion helps as apparently enough salt is transported downward while the surface remains on average slightly cooler than the bottom. Now for both $\Phi_{z,T}$ and $\Phi_{i,T}$ the only time they work against the conversion to either $E_k$ or $E_p$ is during the initial strengthening ($t\approx 4.3$). At that moment there is increased upwelling in the south due to the added freshwater flux, while at the same time the upwelling is inhibited by the cooling of the surface there. Regarding $\Phi_a$ we see that during all stages background energy is added via salinity and removed via heat. This follows from the fact that the freshwater surface forcings always add density (salinity) to the light water away from the downwelling regions while the surface-heating always adds density (cooling) to the heavy water in the downwelling region. The effect is especially pronounced for the two overturning cells. Finally, the stochastic component $\Phi_{a,\Tilde{S}}$ only removes a little background energy by salinifying the southern downwelling region. Striking is that the total energy perturbed by the stochastic component $\left|\int\Phi_{a,\Tilde{S}}\,dt\right|\approx 575$ is only minor compared to the other energy fluxes.

\section{Ratios of collapse probabilities}\label{sec:ratios}
Our bifurcation parameter $\beta$ describes the asymmetry of the freshwater forcing, and hence a different instanton, i.e. most likely transition, and optimal forcing has to be found for every value of $\beta$. While we do not necessarily expect qualitative changes in physical mechanisms mediating the transition, its likelihood will be strongly affected by $\beta$. It is therefore instructive to compare the transition probabilities for different values of the bifurcation parameter.

\begin{figure}
    \centering
    \includegraphics[width=\textwidth]{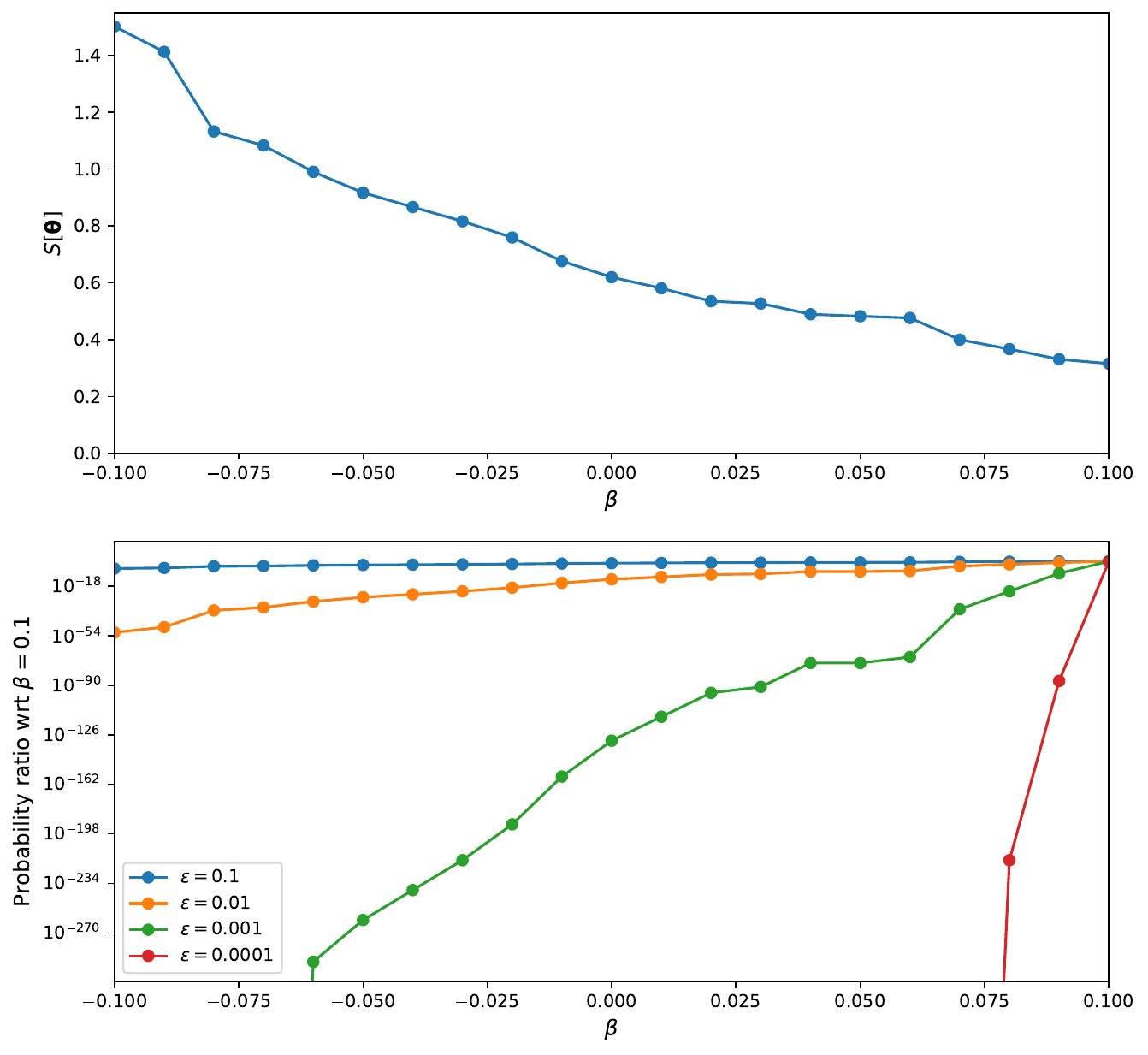}
    \caption{The actions $S[\bt(x,z,t)]$ of the instanton trajectory of the AMOC collapse for several $\beta \in [-0.1,0.1]$ (top), and the resulting probability ratios of these collapses with respect to the collapse under $\beta = 0.1$ for various noise levels $\varepsilon$ (bottom). }
    \label{fig:probratio}
\end{figure}

The leading order term of the ratios of probabilities in the small noise limit can be determined using the Large Deviation Principle (LDP) \eqref{eq:LDP}. Consider two trajectories $\bph_A(x,z,t)$ and $\bph_B(x,z,t)$ obeying the same model \eqref{eq:standardSPDE} with Freidlin-Wentzell actions $S_A$ and $S_B$ respectively, then their relative likelihood is given by
\begin{equation*}
    \frac{\mathbb{P}\left(\bph_B\right)}{\mathbb{P}\left(\bph_A\right)}\asymp \exp\left[(S_A-S_B)/\varepsilon\right].
\end{equation*}
A derivation can be found in the appendix of \citet{soons2023optimal}. Now the actions as well as the ratios of collapse probabilities for various values of bifurcation parameter $\beta$ are shown in figure \ref{fig:probratio} for a range of noise amplitudes $\varepsilon$. The actions are computed as
\begin{equation*}
    S\left[\bt(x,z,t)\right] = \int_0^\tau \left<\bt(x,z,t),\mathcal{A}(\bt)(x,z,t)\right>_{L^2(\Omega)}\,dt.
\end{equation*}
In figure \ref{fig:probratio} one can see a monotone decrease in the action as the deterministic salinity forcing freshens the north of the basin as $\beta$ increases. This can be  expected as more freshwater is added to the southern surface the stochastic salinity forcing has to increase in order to get the surface water sufficiently dense for downwelling. Note that the downward trend of $S[\bt]$ is gradual: there are no qualitative changes in the instanton trajectory as $\beta$ varies and all trajectories show the same dynamics as described in the previous section. 

Regarding the probability ratios we see that the probability of a collapse increases steeply as we approach the bifurcation point. A similar result has been found by \citet{baars2021application} in a comparable model, where the probabilities were numerically computed using a rare-event algorithm. For example, here a small raise in $\beta$ from $0.09$ to $0.1$ increases the probability of collapse with a factor $1.2$, $7.3$, $4.9\cdot10^{8}$ and $7.2\cdot10^{86}$ for the respective noise levels $\varepsilon\in\{10^{-1},10^{-2},10^{-3},10^{-4}\}$, where only the last noise value can be considered realistic. So a small but permanent shift in the large-scale precipitation or evaporation patterns above the Atlantic can alter the probability of a noise-induced AMOC collapse greatly.

\section{Summary \& Discussion}\label{sec:sum}
The most likely paths of a collapse of the overturning circulation in a two-dimensional Boussinesq fluid model for various asymmetric surface salinity forcings were determined. This  model is a member of the hierarchy of models of the Atlantic Meridional Overturning Circulation (AMOC) where the zonally averaged Atlantic Ocean flow \citet{quon1992multiple, thual1992catastrophe} is represented  with constant eddy diffusivities and small stochastic freshwater  surface-forcing. Using Large Deviation Theory the most probable noise-induced transitions from a northern overturning circulation (mimicking an AMOC ON state) to a southern overturning circulation (or  AMOC OFF) state in the low-noise limit were computed. 

Curiously, this collapse starts off with a strengthening of the northern overturning cell. This phenomenon was also found in previous work \citep{soons2023optimal} studying the instanton of an AMOC collapse in the box model by \citet{wood2019observable}. For both models the strengthening puts the system in a less stable transient state by transporting salt quickly to the bottom of the domain. In this way, in the box model  a smaller surface salinity perturbation  is needed to tip the AMOC, whereas in this spatially continuous Boussinesq model the upwelling salinity anomaly is combined with a surface forcing to create one bigger surface perturbation. The net effect is  strong enough vertical salinity gradients to induce a second overturning cell in the south. This two-cell configuration is in energy-imbalance, where the elevated atmospheric energy  input cannot be sufficiently diffused away. This causes a build-up of background potential energy, which is released rapidly as the original northern overturning cell collapses and puts the system into the reversed pole-to-pole circulation. Not only does this method provide these rare transitions, but  it  also allows us to compare probabilities in the low-noise limit between various forcing scenarios. It shows the steep increase in probability of a noise-induced overturning circulation collapse in the low-noise limit as the bifurcation point is approached. 

The advantages of our method are that the problem of finding  these rare stochastic events are transformed into a deterministic optimization problem where no choices regarding surface salinity forcing protocols need to be made. The step forward compared 
to the instantons on box models \citep{soons2023optimal} is that spatial patterns are now identified and that the fluid dynamical mechanisms and energetics of these transitions can be more clearly determined. Obviously, this two-dimensional model is still
a low member in the hierarchy of AMOC models, as the overturning here is solely driven by buoyancy forcing and interior mixing \citep{kuhlbrodt2007driving}, while for the AMOC also wind and tidal forcings are important. Moreover, in order to reproduce the required present-day ocean stratification an open zonal channel in the Southern  Hemisphere is essential 
\citep{henning2005effects, wolfe2010sets}. 

On the other hand, the model does capture the salt-advection feedback well, which is the responsible mechanism for 
the multi-stability of the AMOC \citep{weijer2019stability}. Regarding the stochastic salinity surface-forcing, a number of 
assumptions had to be made. In particular, its spatial spectrum was taken to be uniform up to a cut-off frequency, where the cut-off is heuristically based on annual synoptic precipitation patterns. For its 
temporal structure we chose additive white noise as this is conceptually the most simple and a common assumption in stochastic 
AMOC models \citep{cessi1994simple, timmermann2000noise}. However, the noise may be colored and multiplicative, 
especially under climate change \citep{boot2024}. The Freidlin-Wentzell Theory can be extended to include this type of 
noise \citep{grafke2019numerical}.  The computed instantons are of course only a representable trajectory of the 
noise-induced collapse if the noise is indeed small. As this is the case in other AMOC models \citep{castellana2019transition, soons2023optimal} and since the noise amplitude relative to the AMOC strength should not be depend too much 
on model choice, we can justify this to be the case for the model used here. 

All in all, the merits of this study are the direct computation of a rare overturning collapse in a Boussinesq fluid and the collapse probability ratios between various surface forcings in the low-noise limit. We showed that the Freidlin-Wentzell Theory can also be applied to spatially continuous AMOC models, and a next step would be to apply it to three-dimensional  ocean-only models, such as those used in \citep{dijkstra2007characterization}. Furthermore, Large Deviation Theory can be used to investigate the distinct ways a combined rate- and noise-induced tipping event can occur \citep{slyman2023rate} and it would be interesting to examine this in AMOC models of varying complexity.



\paragraph{Funding} J.S. and H.A.D. are funded by the European Research Council through ERC-AdG project TAOC (project 101055096, PI: Dijkstra). T.G. acknowledges support from EPSRC projects EP/T011866/1 and EP/V013319/1.

\paragraph{Declaration of interests}The authors report no conflict of interest.

\paragraph{Data availability statement}The results can be readily reproduced using the described method and model. For the purpose of open access, the authors have applied a Creative Commons Attribution (CC BY) licence to any Author Accepted Manuscript version arising from this submission. 


\paragraph{Author ORCIDs} J.Soons, https://orcid.org/0009-0007-6635-5057


\newpage
\appendix

\section{The Variational Derivatives and their Adjoints}\label{appA}
\subsection{Variational Derivatives of the Deterministic Drift}
Given is the deterministic drift function
\begin{align*}
    f_1[\bph] &= \partial_x\psi[\omega]\partial_z\omega - \partial_z\psi[\omega]\partial_x\omega +\Pran\nabla^2\omega+\Pran\Ray\left(\partial_xT-\partial_xS\right)\\
    f_2[\bph] &= \partial_x\psi[\omega]\partial_zT - \partial_z\psi[\omega]\partial_xT + \nabla^2T + \frac{h(z)}{\tau_T}\left(T_S(x)-T\right)\\
    f_3[\bph] &= \partial_x\psi[\omega]\partial_zS - \partial_z\psi[\omega]\partial_xS + \Lew^{-1}\nabla^2S + \frac{h(z)}{\tau_S}S_S(x)
\end{align*}
where $\bph:\,\Omega\times\R_{\geq0}\to\R^3$ with $\bph(x,z,t) = (\omega, T, S)^T(x,z,t)$ and spatial boundary conditions as in section \ref{sec:model}. The variational derivatives with respect to $\omega$ acting on a function $v_1:\,\Omega\times\R_{\geq0}\to\R$ are
\begin{align*}
    D_\omega f_1[\bph](v_1) &= \partial_x\psi[\omega]\partial_zv_1+\partial_x\mathcal{G}(v_1)\partial_z\omega -\partial_z\psi[\omega]\partial_zv_1-\partial_z\mathcal{G}(v_1)\partial_x\omega+\Pran\nabla^2v_1\\
    D_\omega f_2[\bph](v_1) &= \partial_x\mathcal{G}(v_1)\partial_zT - \partial_z\mathcal{G}(v_1)\partial_xT\\
    D_\omega f_3[\bph](v_1) &= \partial_x\mathcal{G}(v_1)\partial_zS - \partial_z\mathcal{G}(v_1)\partial_xS
\end{align*}
where we have boundary condition $v_1(x,z,t) = 0$ for $(x,z)\in\partial\Omega$. The variational derivatives with respect to $T$ acting on a function $v_2:\,\Omega\times\R_{\geq0}\to\R$ are
\begin{align*}
    D_T f_1[\bph](v_2) &= \Pran\Ray\partial_xv_2\\
    D_T f_2[\bph](v_2) &= \partial_x\psi[\omega]\partial_zv_2-\partial_z\psi[\omega]\partial_xv_2+\nabla^2v_2-\frac{h(z)}{\tau_T}v_2\\
    D_T f_3[\bph](v_2) &= 0
\end{align*}
where we have boundary conditions $\partial_xv_2(x,z,t) = 0$ for $x = 0$ or $x = \Asp$, and $\partial_zv_2(x,z,t) = 0$ for $z = 0$ or $z = 1$. The variational derivatives with respect to $S$ acting on a function $v_3:\,\Omega\times\R_{\geq0}\to\R$ are
\begin{align*}
    D_S f_1[\bph](v_3) &= -\Pran\Ray\partial_xv_3\\
    D_S f_2[\bph](v_3) &= 0\\
    D_S f_3[\bph](v_3) &= \partial_x\psi[\omega]\partial_zv_3-\partial_z\psi[\omega]\partial_xv_3+\Lew^{-1}\nabla^2v_3
\end{align*}
where we have boundary conditions $\partial_xv_3(x,z,t) = 0$ for $x = 0$ or $x = \Asp$, and $\partial_zv_3(x,z,t) = 0$ for $z = 0$ or $z = 1$.

\subsection{The Adjoints of the Variational Derivatives}
The adjoints of the derivatives of the first component acting on a function $u_1:\,\Omega\times\R_{\geq0}\to\R$ are
\begin{align*}
    \left(D_\omega f_1[\bph]\right)^*(u_1) &= \partial_z\psi[\omega]\partial_xu_1-\partial_x\partial_zu_1-\mathcal{G}\left(\partial_xu_1\partial_z\omega-\partial_zu_1\partial_x\omega\right)+\Pran\nabla^2u_1\\
    \left(D_T f_1[\bph]\right)^*(u_1) &=-\Pran\Ray\partial_xu_1\\
    \left(D_S f_1[\bph]\right)^*(u_1) &=\Pran\Ray\partial_xu_1
\end{align*}
where we have boundary condition  $u_1(x,z,t) = 0$ for $(x,z)\in\partial\Omega$. The adjoints of the derivatives of the second component acting on a function $u_2:\,\Omega\times\R_{\geq0}\to\R$ are
\begin{align*}
    \left(D_\omega f_2[\bph]\right)^*(u_2) &= \mathcal{G}\left(\partial_zu_2\partial_xT-\partial_xu_2\partial_zT\right)\\
    \left(D_T f_2[\bph]\right)^*(u_2) &= \partial_z\psi[\omega]\partial_xu_2-\partial_x\psi[\omega]\partial_zu_2+\nabla^2u_2-\frac{h(z)}{\tau_t}u_2\\
    \left(D_S f_2[\bph]\right)^*(u_2) &= 0
\end{align*}
where we have boundary conditions $\partial_xu_2(x,z,t) = 0$ for $x = 0$ or $x = \Asp$, and $\partial_zu_2(x,z,t) = 0$ for $z = 0$ or $z = 1$. The adjoints of the derivatives of the third component acting on a function $u_3:\,\Omega\times\R_{\geq0}\to\R$ are
\begin{align*}
    \left(D_\omega f_3[\bph]\right)^*(u_3) &= \mathcal{G}\left(\partial_zu_3\partial_xS-\partial_xu_3\partial_zS\right)\\
    \left(D_T f_3[\bph]\right)^*(u_3) &= 0\\
    \left(D_S f_3[\bph]\right)^*(u_3) &= \partial_z\psi[\omega]\partial_xu_3-\partial_x\psi[\omega]\partial_zu_3+\Lew^{-1}\nabla^2u_3
\end{align*}
where we have boundary conditions $\partial_xu_3(x,z,t) = 0$ for $x = 0$ or $x = \Asp$, and $\partial_zu_3(x,z,t) = 0$ for $z = 0$ or $z = 1$.

\section{The Augmented Lagrangian Method}\label{appB}
Using the Augmented Lagrangian Method (ALM) equations \eqref{eq:phi} and \eqref{eq:theta} are solved. We define the cost function:
\begin{align*}
    &J[\bph(x,z,t),\theta_S(x,z,t),\boldsymbol{\mu}(x,z,t),\boldsymbol{\gamma},\lambda]\\
    =& \int_0^\tau\int_\Omega\frac{(h(z))^2}{2K\tau_S^2}\theta_S\mathcal{U}^*\mathcal{P}\mathcal{U}(\theta_S)\,dx\,dz\,dt + \int_0^\tau\int_\Omega \bigg(\mu_\omega\left(\partial_t\omega-f_1(\bph)\right)\\
    &+ \mu_T\left(\partial_tT-f_3(\bph)\right)+ \mu_S\left(\partial_tS-f_2(\bph) - \frac{(h(z))^2}{K\tau_S^2}\mathcal{U}^*\mathcal{P}\mathcal{U}(\theta_S)\right) \bigg)\,dx\,dz\,dt\\
    &+ \lambda \langle\bph(x,z,\tau),\bph_{\OFF}(x,z)\rangle_{L^2} + \langle\boldsymbol{\gamma}, \bph(x,z,\tau)-\bph_{\OFF}(x,z)\rangle_{L^2}
\end{align*}
where $\boldsymbol{\mu}:\,\Omega\times\R_{\geq0}\to\R^3$ with $\boldsymbol{\mu}(x,z,t) = \left(\mu_\omega,\mu_T,\mu_S\right)^T(x,z,t)$ is our control variable and $\lambda\in\R_{\geq0}$ and $\boldsymbol{\gamma}:\,\Omega\to\R^3$ with $\boldsymbol{\gamma}(x,z) = \left(\gamma_\omega, \gamma_T,\gamma_S\right)^T(x,z)$ are penalty parameters to enforce the end conditions. The conjugate momenta $\theta_\omega$ and $\theta_T$ are omitted as there is no direct stochastic forcing onto the vorticity nor the temperature, and so no cost is assigned to them. Now, minimizing the cost function is equivalent to solving the instanton equations since we have
\begin{align*}
    \partial_{\boldsymbol{\phi}}J = 0 &\iff \partial_t\boldsymbol{\mu} = -\left(\nabla_{\boldsymbol{\phi}}f\right)^*\boldsymbol{\mu}\\
    \partial_{\boldsymbol{\mu}}J = 0 &\iff \partial_t\bph = f(\bph) + (h(z))^2\big/K\left(0,\,0,\,\mathcal{U}^*\mathcal{P}\mathcal{U}(\theta_S)\right)^T\\
    \partial_{\theta_S}J = 0 &\iff \theta_S = \mu_S\\
    \partial_{\boldsymbol{\gamma}}J = 0 &\iff \bph(x,z,\tau) = \bph_{\OFF}(x,z)\\
    \partial_{\lambda}J = 0 &\iff \bph(x,z,\tau) = \bph_{\OFF}(x,z)
\end{align*}
together with the initial condition $\bph(x,z,0) = \bph_{\ON}(x,z)$. The end condition for the control variable is found by
\begin{equation*}
    \partial_{\boldsymbol{\mu}(x,z,\tau)}J = 0 \iff \boldsymbol{\mu}(x,z,\tau) = -\left(2\lambda\left(\bph(x,z,\tau)-\bph_{\OFF}\right)+\boldsymbol{\gamma}\right).
\end{equation*}
In order to minimize this cost function, and hence find the minimizing arguments i.e. the instanton $\boldsymbol{\Tilde{\phi}}$ and its associated forcing $\Tilde{\theta}_S$, we employ the following protocol. Firstly, we pick the penalty parameters $\lambda$ and $\boldsymbol{\gamma}$ and then reiterate the scheme:
\begin{enumerate}
    \item forward integration
    \begin{align*}
        \begin{cases}
            &\partial_t\bph = f(\bph) +  \frac{(h(z))^2}{\tau_S^2K}\left(0,\,0,\,\mathcal{U}^*\mathcal{P}\mathcal{U}(\theta_S)\right)^T\\
            &\bph(x,z,0) =\bph_{\ON}(x,z)
        \end{cases}
    \end{align*}
    \item backward integration 
    \begin{align*}
        \begin{cases}
            &\partial_t\boldsymbol{\mu} = -\left(\nabla_{\boldsymbol{\phi}}f\right)^*\boldsymbol{\mu}\\
            &\boldsymbol{\mu}(x,z,\tau) = -\left(2\lambda\left(\bph(x,z,\tau)-\bph_{\OFF}(x,z)\right)+\boldsymbol{\gamma}(x,z)\right)
        \end{cases}
    \end{align*}
    \item gradient computation
    \begin{equation*}
        \partial_{\theta_S}J = \theta_S - \mu_S
    \end{equation*}
    \item updating the conjugate momentum 
    \begin{equation*}
        \theta_S^{\text{new}} = \theta_S + \alpha \partial_{\theta_S}J
    \end{equation*}
\end{enumerate}   
with step size $\alpha\in\R_{>0}$ such that $J[\bph^{\text{new}}, \theta_S^{\text{new}}, \boldsymbol{\mu}, \boldsymbol{\gamma},\lambda] < J[\bph, \theta_S, \boldsymbol{\mu}, \boldsymbol{\gamma},\lambda]$ where $\bph^{\text{new}}$ is the trajectory forced by $\theta_S^{\text{new}}$. This is repeated until it can be concluded that $J$ is minimized for the given $\lambda$ and $\boldsymbol{\gamma}$, which can be indicated by for example $\|\partial_{\theta_S}J\|_{L^2}$ being sufficiently small. These penalty parameters are then updated as
\begin{enumerate}
    \item $\boldsymbol{\gamma}^{\text{new}}(x,z) = -\mu(x,z,\tau)$
    \item $\lambda^{\text{new}} = f_\lambda\lambda$ with a constant parameter $f_\lambda > 1$
\end{enumerate}
and then the first scheme is reiterated again. This process is repeated until $\bph(x,z,\tau)$ is sufficiently close to $\bph_{\OFF}$.

The instanton is now computed as follows. A trajectory is computed using ALM, and this trajectory is relaxed as early as possible. This is done as ALM converges slowly to the needed relaxation after the trajectory has crossed the separatrix. Then again ALM is used, but now to a point on the relaxed trajectory after the separatrix instead of running it to $\bph_{\OFF}$. This newly computed trajectory can now continue unforced till the collapsed state $\bph_{\OFF}$. This yields the instanton that we will use. The advantage here is that we now only optimize the initial forced part of the trajectory using ALM. For the instantons in this work $\tau = 10$ was sufficient for the duration of the forced part. Any higher $\tau$ results in a trajectory with the same dynamical behaviour, but just shifted in time where the additional time is simply spent in the initial ON state.

Our termination conditions for the above mentioned iterations are as follows. The minimization for a given $\lambda$ and $\boldsymbol{\gamma}$ is terminated when the cost function has decreases by less than $1\%$, and the whole algorithm is terminated when $\bph(x,z,\tau)$ is sufficiently close to $\bph_{\text{end}}$. This is defined by
\begin{equation*}
    \underset{(x,z)\in\Omega}{\text{max}}\|\phi_i(x,z,\tau)-\phi_{\text{end},i}(x,z)\|\bigg/\overline{\phi_{\text{end},i}} < 10^{-3}\qquad\text{for all }i\in\{1,2,3\}
\end{equation*}
where $\overline{\phi_i}$ indicates the spatial mean of $\phi_i$.

\bibliographystyle{jfm}
\bibliography{references}

\end{document}